\documentclass[a4paper,12pt]{article}

\usepackage[utf8]{inputenc}
\usepackage{mathptmx}
\usepackage[margin=1in]{geometry}
\usepackage{amsfonts}
\usepackage{amsmath, amsthm}
\usepackage{amssymb}
\usepackage{tikz}
\usetikzlibrary{arrows.meta}
\usepackage{graphicx}
\usepackage{natbib}
\usepackage{appendix}
\usepackage{subcaption}
\usepackage{bbm}
\usepackage{color}
\usepackage{url}
\usepackage[pdftex,colorlinks=true,linkcolor=blue,citecolor=blue,urlcolor=blue]{hyperref}
\usepackage{tabularx}
\usepackage{caption}
\usepackage{multirow}
\usepackage{booktabs}
\usepackage[onehalfspacing]{setspace}

\newtheorem{assumption}{Assumption}

\newtheorem{proposition}{Proposition}
\newtheorem{corollary}{Corollary}
\newtheorem{remark}{Remark}

\begin{document}

\title{Causal Inference under Kink Bunching\thanks{Yi Lu, Tsinghua University, luyi@pbcsf.tsinghua.edu.cn; Jianguo Wang, Renmin University of China, wangjianguo@rmbs.ruc.edu.cn; Huihua Xie, Zhejiang University, huihuaxie@zju.edu.cn. We thank Jérôme Adda, Songnian Chen, Thomas Cornelissen, Silvia Goncalves, Erzo F.P. Luttmer, Valentina Melentyeva, Alexandra Spitz-Oener, Zhuan Pei, Andrea Weber, Daniel Yi Xu, Marinho Bertanha, Alisa Tazhitdinova, as well as seminar and conference participants at the 2024 North American Meetings of the Econometric Society, the 2024 Asia Meetings of the Econometric Society, the 2024 RFBerlin-CReAM Workshop, the 2025 AEA Annual Meeting, Renmin University of China, Shanghai University of Finance and Economics, Southwest University of Finance and Economics, Zhejiang University for their valuable comments. All errors are our own.}}
\author{Yi Lu  \and Jianguo Wang  \and Huihua Xie}
\maketitle

\vspace{-30pt}
\begin{abstract}

Kinked policies change marginal incentives at a threshold, and agents respond by adjusting the assignment variable that determines their treatment. The bunching literature uses this response to estimate the elasticity of the assignment variable; policymakers often care about effects on other outcomes. We develop a framework for estimating causal effects of kinked policies on outcomes beyond the assignment variable when agents can fully manipulate it. Average effects are defined for two affected populations: bunchers, who locate at the kink, and shifters, who reduce their assignment values but remain above the threshold; shifter effects compare equal-mass intervals and require only rank invariance. Because a single kink identifies neither the counterfactual assignment density nor the counterfactual outcome function, identification is design-assisted: placebo groups and moving thresholds discipline the local shape of both objects, the focal group's unaffected observations pin down level and slope differences, and the restrictions are testable. Applying the framework to a kinked coinsurance schedule in China's medical insurance, we find that the loss of reimbursement above the annual cap sharply reduces outpatient visits, raises cost per visit, and shifts the composition of care toward hospitals---effects that are invisible in a density-only bunching analysis.


\end{abstract}



\thispagestyle{empty} 

\clearpage 
\pagenumbering{arabic} 
\setcounter{page}{1} 
\section{Introduction}

Many public policies change marginal incentives sharply at a threshold. Income tax schedules raise marginal rates above bracket cutoffs; health insurance plans change cost sharing once annual spending crosses a limit; social insurance, education, housing, and firm-subsidy programs attach different marginal incentives to different regions of an assignment variable. Agents respond by adjusting the very variable that determines their treatment: taxpayers reduce taxable income, patients curtail eligible medical spending, and firms adjust reported investment or size.

The bunching literature has turned these responses into a powerful measurement tool. Excess buncing in the density of the assignment variable at a kink reveals how strongly agents respond to the change in marginal incentives, and a large body of work uses it to estimate response elasticities in taxation, health care, housing, and firm behavior \citep{saez2010taxpayers,chetty2011adjustment,kleven2013using,einav2015response,best2020estimating}; see \citet{kleven2016bunching} and \citet{bertanha2022bunching} for reviews. The object of this literature, however, is mianly about the response of the assignment variable itself. 

Yet the assignment variable is rarely the outcome that policymakers ultimately care about. A reimbursement cap in health insurance is not evaluated by how much eligible spending bunches at the cap, but by whether patients forgo care once coverage is exhausted. A kinked tax schedule matters not only for reported income, but for how tax-induced income responses feed into household health spending or children's education. 
These are questions about a downstream outcome $Y$, observed alongside the assignment variable $X$ but distinct from it.

This paper develops a framework for identifying and estimating causal effects of kinked policies on outcomes other than the assignment variable, in precisely the settings the bunching literature studies: agents can fully adjust the assignment variable. The problem combines the difficult halves of two familiar designs. Regression kink designs target effects on outcomes, but their validity requires that agents do not precisely manipulate the running variable \citep{card2017regression}; bunching designs are built around manipulation, but typically stop at the assignment variable. Here the outcome of interest is not the assignment variable, and the assignment variable is itself endogenous to the policy. The aim is to turn kink bunching from a method for estimating responses of $X$ into a design for estimating causal effects on $Y$.

Direct comparisons fail in this setting. Under the kinked policy, the observed conditional mean $\mathbb{E}[Y^{kp}\mid X^{kp}=x]$ is contaminated twice over: the kink moves assignment values, so the agents observed at $x$ are generally not those who would have chosen $x$ absent the kink, and the kink changes the policy transfer associated with any given assignment value. Our framework therefore indexes agents by their \emph{latent counterfactual assignment} $X^{cf}$---the value they would choose under the counterfactual linear schedule. The kink partitions agents into three groups: agents below the threshold are unaffected; agents who would have located within $(x^*,x^*+\Delta x^*]$ collapse to the threshold (\emph{bunchers}); and agents from higher up reduce their assignment values but remain above the threshold (\emph{shifters}). The estimands are average \emph{total} effects of the kinked policy on $Y$ for bunchers and for shifters in an observed evaluation window, $\tau_B$ and $\tau_S$, inclusive of both the assignment-variable and the transfer channel. A deliberate feature of the framework is how little behavioral structure these estimands require: rank invariance among shifters---an implication of the canonical bunching model, but far weaker than its  functional form---suffices to pair any observed window of shifters with a counterfactual interval of equal mass.  
No pointwise relocation rule assigning each shifter to an exact counterfactual value is needed.

Identification  faces two layers of missing information, and we are explicit that a single kink supplies neither. On the assignment side, the counterfactual density $h_F^{cf}$ and the marginal bunching response $\Delta x^*$ are not separately identified from one kink with an unrestricted counterfactual density---the non-identification result emphasized by \citet{blomquist2021bunching} and \citet{bertanha2023better}. On the outcome side, a further problem arises that has no analogue in density-only bunching studies: even if every shifter could be assigned her counterfactual location, treated data reveal only the sum 
of the counterfactual conditional outcome mean function and the treatment effect, so the two are not separable without additional information. Estimating effects on $Y$ therefore requires disciplining both the counterfactual assignment density and the counterfactual outcome function around the kink. 

Our identification is design-assisted: we use policy variation across groups and over time, relatively common in bunching settings. A placebo group not subject to the focal kink in a local window---agent with a different threshold, or the same population in a year when the threshold sat elsewhere---reveals the local shape of the density and of the outcome function absent the kink. The placebo group is not required to match the focal group's levels or slopes. Instead, after allowing the two groups to differ by a low-dimensional component (level and slope in the baseline), the remaining local shape is assumed common, and this component is identified from the focal group's own unaffected observations below the threshold, where the kinked and counterfactual policies coincide. The restriction is analogous to a difference-in-differences condition applied to local \emph{shapes} of density and outcome functions rather than to time trends: what it rules out is a focal-specific bump, bend, or curvature change in the missing region that is absent from the placebo group. The behavioral structure and the placebo group are complements: 
Without the bunching structure---mass balance and rank invariance---the placebo group cannot say where bunchers and shifters originate; 
without the placebo group, the counterfactual shapes rest on functional form alone. Importantly, the identifying restrictions have testable implications: the residualized focal--placebo differences should be flat below the threshold, the procedure should detect nothing at placebo thresholds, and estimates should be stable across alternative placebo groups. Estimation is a transparent plug-in on binned data, with bootstrap inference over the entire pipeline.\footnote{When no credible placebo variation exists, point identification should not be claimed from a single kink; the same framework then delivers partial-identification bounds under explicit shape restrictions.}

The data requirements are deliberately modest: the inputs of a standard bunching study, an outcome measured alongside the assignment variable, and a placebo group or a threshold that moves. Because policy thresholds are routinely updated over time and parallel programs with different thresholds routinely coexist, we expect the design to be widely applicable.

We apply the framework to the kinked outpatient coinsurance schedule of China's Urban and Rural Residents Basic Medical Insurance (URRBMI). Below an annual cap on eligible outpatient expenditure, enrollees pay a coinsurance rate of $50$--$60\%$; above the cap, reimbursement stops and they bear the full marginal cost---a convex kink at the cap. The setting supplies the design ingredients: the cap moved from $600$ to $800$~RMB between 2011 and 2012, and the parallel Urban Employee program (UEBMI) has far higher caps, so its enrollees serve as a placebo group around the URRBMI thresholds. Using the universe of outpatient claims from a prefecture in eastern China---about $40$ million URRBMI visits over 2011--2012, plus the corresponding UEBMI claims---we document sharp bunching in eligible annual expenditure exactly at the year-specific cap; the excess mass tracks the cap as it moves, and the UEBMI density is smooth through both pseudo thresholds. Because the assignment variable is built from claims filed directly by providers, the bunching reflects real adjustment of care rather than relabeling or misreporting. The design-assisted estimates imply a marginal bunching response of $\Delta x^*=174$~RMB (bootstrap s.e.\ $3.1$), meaning the marginal buncher gives up about $18\%$ of her counterfactual eligible spending. The effects on utilization are large: shifters---enrollees who would have spent well past the cap---cut annual outpatient visits from a counterfactual $15.7$ to an observed $6.7$, and bunchers cut visits by $2.6$; average cost per visit rises by roughly $55\%$ for shifters, the composition of care tilts from community health centers toward hospitals, and capped enrollees stop seeking care up to a month earlier in the year. The picture is of capped enrollees forgoing routine, low-cost care and reserving out-of-pocket spending for serious conditions. The identifying restrictions survive their placebo tests: at the lapsed $600$~RMB threshold in 2012 the estimated marginal response is $1$~RMB and the estimated effects are statistically zero. These utilization consequences---the objects of policy interest---are invisible in a density-only bunching analysis.

A set of extensions shows how the framework adapts when baseline conditions are weakened, each modifying one module while the rest of the pipeline is unchanged: heterogeneity in responsiveness (condition on covariates, with cell-specific $\Delta x^*(w)$ as a diagnostic), optimization frictions and stayers, reference-point and round-number bunching (the placebo group shares the rounding environment, so the correction is disciplined rather than assumed), relabeling and misreporting, alternative counterfactual policies, and structural relocation maps that deliver pointwise effects and channel decompositions for researchers willing to impose more.

This paper contributes to three literatures. First, it contributes to the bunching literature \citep{saez2010taxpayers,chetty2011adjustment,kleven2013using,kleven2016bunching}, including the recent econometric work on its identifying content \citep{blomquist2021bunching,bertanha2023better,cox2021market,anagol2024diffuse,pollinger2025kinks}. We shift the object of interest from the response of the assignment variable to causal effects on downstream outcomes, using the bunching response as a behavioral first stage. In doing so we also contribute to the construction of bunching counterfactuals: rather than extrapolating a polynomial fitted to the treated distribution alone \citep{chetty2011adjustment}, the counterfactual density---and, new here, the counterfactual outcome function---are disciplined by an observable placebo group, which addresses the single-kink non-identification critique with an untestable functional-form choice with restrictions that have testable implications.

Second, it contributes to the emerging literature on treatment effects in bunching designs. \citet{diamond2017long} study a notched setting in which manipulation is local to the threshold, so counterfactual outcome for agents inside the manipulation window can be predicted from observations outside the window.
 \citet{caetano2015test}, \citet{caetano2023correcting}, and \citet{caetano2022identification,caetano2025identification} exploit bunching generated by corners or censoring in a treatment variable to detect and correct endogeneity and to identify marginal treatment effects. \citet{goff2022treatment} studies kinks under overtime regulation, where the outcome of interest is the bunching variable itself. Our setting is complementary and, in a specific sense, harder: at a kink, \emph{every} agent on the upper side of the threshold has an incentive to adjust, so there is no unmanipulated window above the kink, and the outcome is distinct from the assignment variable. The framework handles exactly this configuration---interior responses everywhere above the threshold, with aggregate estimands that require rank invariance rather than a pointwise behavioral model.

Third, the paper contributes to the quasi-experimental toolkit for policies with nonlinear incentives. Regression discontinuity and regression kink designs identify effects on outcomes at thresholds but are threatened by precise manipulation of the running variable \citep{lee2010regression,card2017regression}; bunching designs embrace manipulation but have largely confined attention to the manipulated variable. Our framework connects the two: it recovers causal effects on outcomes in settings where manipulation is the central behavioral response to the policy. 

The rest of the paper proceeds as follows. Section~\ref{sec:framework} defines the policy environment, the behavioral groups, and the treatment-effect estimands. Section~\ref{sec:assignment-identification} identifies the counterfactual assignment distribution and the marginal bunching response; Section~\ref{sec:outcome-identification} identifies the counterfactual outcome function and the treatment effects. Section~\ref{sec:estimation} presents estimation and inference. Section~\ref{sec:application} applies the framework to China's kinked coinsurance schedule. Section~\ref{sec:extensions} develops extensions and sensitivity analyses, and Section~\ref{sec:conclusion} concludes.

\section{Setup, Counterfactual Objects, and Estimands}
\label{sec:framework}

This section defines the causal objects studied in the paper. The central difficulty is that the assignment variable is itself affected by the kinked policy.
As a result, the observed conditional outcome function, $\mathbb E[Y^{kp}\mid X^{kp}=x,F],$ does not describe the outcome that would have been observed at assignment value \(x\) absent the kink.
 Agents observed at a given value  are generally not the agents who would have chosen that value under the counterfactual linear policy.

The framework indexes populations by their latent counterfactual assignment values. Let \(X^{cf}\) denote the assignment value an agent would choose under a counterfactual linear policy, and let \(X^{kp}\) denote the assignment value observed under the kinked policy. The main estimands below are average effects for two affected populations: bunchers, whose observed mass appears as excess bunching around the cutoff, and shifters, who remain above the cutoff under the kinked policy. 

\subsection{Policy Environment}
\label{sec:policy-environment}

Consider a focal group \(F\) subject to a kinked policy. Let \(X\) denote the assignment variable, such as taxable income, eligible medical spending, investment, or another variable that determines the marginal incentive faced by the agent. The policy has a threshold \(x^*\). Below the threshold, agents face marginal rate \(t\). Above the threshold, agents face marginal rate \(t+\Delta t\), where \(\Delta t>0\). The payment schedule under the kinked policy is
\begin{equation}
T^{kp}(x)=
\begin{cases}
tx, & x\leq x^*,\\
tx+\Delta t(x-x^*), & x>x^*.
\end{cases}
\end{equation}
The counterfactual policy is a linear schedule with the below-threshold marginal rate: 
\begin{equation}
T^{cf}(x)=tx.
\end{equation}
This counterfactual policy is the benchmark relative to which treatment effects are defined. It asks what assignments and outcomes would have been observed if agents had faced the same marginal incentive above the threshold as they face below the kink.

\begin{figure}[t]
\centering
\begin{tikzpicture}[scale=0.85, >=Latex]

    \draw[->, thick] (0,0) -- (6.2,0) node[below] {$x$};
    \draw[->, thick] (0,0) -- (0,4.2) node[left] {$T(x)$};

    \draw[dashed, gray] (2.6,0) -- (2.6,3.7);
    \node[below] at (2.6,0) {$x^*$};

    \draw[thick, dashed, blue] (0,0) -- (5.8,2.9)
        node[right] {$T^{cf}(x)=tx$};

    \draw[thick, red] (0,0) -- (2.6,1.3);
    \draw[thick, red] (2.6,1.3) -- (5.8,3.85)
        node[right] {$T^{kp}(x)$};

    \node[blue] at (1.35,0.9) {$t$};
    \node[red] at (4.25,3.1) {$t+\Delta t$};

    \filldraw[red] (2.6,1.3) circle (1.5pt);
    \node[above left] at (2.6,1.3) {kink};

    \node[align=center] at (1.25,-0.55) {below };
    \node[align=center] at (4.45,-0.55) {above };

\end{tikzpicture}
\caption{Kinked policy schedule and linear counterfactual. The kinked policy coincides with the counterfactual linear schedule below $x^*$ and changes the marginal rate above the cutoff.}
\label{fig:kinked-policy-schedule}
\end{figure}

\subsection{Potential Assignments and Behavioral Groups}
\label{sec:potential-assignments}

Each agent is indexed by a latent type \(\eta\), which captures preferences, productivity, health status, or other determinants of assignment choices and outcomes. Agents choose their assignment value by maximizing utility subject to the policy schedule in Section~\ref{sec:policy-environment}.

To connect with the standard bunching model, recall that the canonical setting in \citet{saez2010taxpayers} implies the assignment equation:\footnote{For example, in the quasi-linear taxable-income version of the model, an agent solves
\[
\max_x\ c-\frac{\eta}{1+1/e}\left(\frac{x}{\eta}\right)^{1+1/e},
\qquad  \text{subject to:}
c=(1-\tilde t)x,
\]
which gives the first-order condition 
\(x(\eta,\tilde t)=\eta(1-\tilde t)^e\).} 
\(  x(\eta,\tilde t)=\eta(1-\tilde t)^e, \label{eq:saez-assignment}\)
where $\tilde{t}$ is marginal tax rate, and $e$ is the elasticity parameter.
It delivers two familiar implications: (i) assignment choices are increasing in the scalar type, and (ii) a higher marginal rate reduces the assignment choice. The main analysis uses these implications directly, while relaxing the exact functional form. 

\begin{assumption}[Monotone assignment choice with scalar heterogeneity]\label{assume:monotone} 
The scalar type \(\eta\) is the only source of heterogeneity in the assignment choice \(x(\eta,\tilde t)\).\footnote{Section~\ref{sec:heterogeneity-covariates} relaxes the scalarity restriction by allowing the responsiveness (elasticity) to differ across subgroups.} Moreover,
\begin{itemize}
    \item[(a)] for each $\tilde t\in\{t,\,t+\Delta t\}$, $x(\eta,\tilde t)$ is continuous and strictly increasing in $\eta$;  
    \item[(b)] for each $\eta$, $x(\eta,t+\Delta t)< x(\eta,t)$. 
\end{itemize}
\end{assumption}
\noindent %
The assumption embeds the no-income-effects convention standard in the bunching literature since \citet{saez2010taxpayers}: locally, the assignment choice depends on the marginal rate faced at the optimum.\footnote{With income effects, all shifters above \(x^*\) face the same schedule segment,  with identical virtual income $\Delta t\,x^*$.}  


 Let $X^{cf}(\eta)=x(\eta, t)$ denote the assignment value agent \(\eta\) would choose under the counterfactual linear policy, and let $X^{kp}(\eta)$ denote the assignment value chosen under the kinked policy. Under Assumption~\ref{assume:monotone}, the kink partitions agents into three groups. 

\begin{itemize}
\item First, agents with counterfactual assignments below the threshold, $X^{cf}(\eta)\leq x^*,$ face the same marginal rate under the kinked and counterfactual policies. Their assignment choices are unchanged: $X^{kp}(\eta)=X^{cf}(\eta).$ We call them \textit{unaffected agents}, denoted by \(U\).

\item Second, agents whose counterfactual assignments lie just above the threshold reduce their assignment values and move to the kink. Consider the marginal bunching agent with $\eta_H$ who sets $X^{kp}(\eta_H)=xs(\eta_H, t+\Delta t)=x^*$ and $X^{cf}(\eta_H)=x(\eta_H, t)=x^*+\Delta x^*$, where $\Delta x^*$ is the marginal bunching agent's response. Hence, agents with $X^{cf}(\eta)\in (x^*,x^*+\Delta x^*]$ move to the threshold: $x^{kp}(\eta)=x^*$, denoted as \textit{bunchers}. The counterfactual assignment interval for bunchers is  $I_B^{cf}=(x^*,x^*+\Delta x^*].$

\item Third, agents with sufficiently high counterfactual assignments respond to the higher marginal rate by reducing their assignment value but remain above the threshold. These agents satisfy
$ X^{cf}(\eta)>x^*+\Delta x^*$, and $x^{kp}(\eta)>x^*$.
We call them \textit{shifters}, denoted by \(S\).

 Further, Assumption \ref{assume:monotone} implies: 
$ X^{kp}=x\bigl(x^{-1}(X^{cf};t),\,t+\Delta t\bigr)\equiv R(X^{cf};t,t+\Delta t), \forall X^{cf}>x^*+\Delta x^*,
$
Thus, $X^{cf}(\eta_1)<X^{cf}(\eta_2)$ implies $X^{kp}(\eta_1)<X^{kp}(\eta_2)$ among shifters.  



\end{itemize}

Combining the three cases yields the following property: 

\begin{corollary}[Rank invariance]\label{cor:rank-invariance}
Under Assumption~\ref{assume:monotone}, the kinked-policy assignment is a common, weakly increasing transformation of the counterfactual assignment,
\begin{equation}\label{eq:xkp-xcf}
X^{kp}=\bar R\!\left(X^{cf}\right)=
\begin{cases}
X^{cf},
& X^{cf}\leq x^*,\\[0.4em]
x^*,
& X^{cf} \in (x^*,x^*+\Delta x^*],\\[0.4em]
 R(X^{cf}; t, t+\Delta t),
&X^{cf}>x^*+\Delta x^*,
\end{cases}
\end{equation}
where $R$ is strictly increasing above the buncher interval. 
The rank ordering of shifters is preserved across the counterfactual and kinked regimes; bunchers collapse to the  threshold \(x^*\).  

\end{corollary}

\begin{remark}[Constant-elasticity microfoundation]\label{rmk:constant-elasticity}
In \citet{saez2010taxpayers}'s canonical bunching model where $x(\eta,\tilde t)=\eta(1-\tilde t)^{e}$, the shifter map is proportional:  
\(   X^{kp}=X^{cf}\left(\frac{1-t-\Delta t}{1-t}\right)^{e}=X^{cf}\,\frac{x^*}{x^*+\Delta x^*},\forall X^{cf}>x^*+\Delta x^*.\) 
The aggregate estimands developed below require only rank-invariance property (Corollary~\ref{cor:rank-invariance}), not the porportional pointwise relocation rule implied in \citet{saez2010taxpayers}.\footnote{A parametric pointwise relocation rule is needed only for pointwise treatment-effect functions or channel decompositions that require assigning each observed shifter to an exact counterfactual value; we discuss those stronger structural exercises in Section~\ref{sec:structural-models}.}  
\end{remark} 

Denote the observed density of the assignment variable under the kinked policy as $h^{kp}(x)$, and the unobserved undelying counterfactual density as $h^{cf}(x)$. Let $B$ denote the excess bunching mass at the cutoff, with $B=h^{kp}(x^*)-h^{cf}(x^*)$, then, the mass-balance condition implies 
\begin{equation}\label{eq:B-M}
B=\int_{x^*+\epsilon}^{x^*+\Delta x^*} h^{cf}(x) dx.
\end{equation}

The same logic applies to shifters: under monotonicity, an observed window of shifters above $x^*$ maps to a counterfactual interval above $x^*+\Delta x^*$ of equal mass. 
For example, for shifters with the observed  interval as  $X^{kp} \in (x^*,x^{kp,out}],$  the counterfactual interval for the same population is  $ X^{cf} \in (x^*+\Delta x^*, x^{cf,out}]$, where the upper endpoint \(x^{cf,out}\) is defined by the equal-mass condition:\footnote{ 
If \(h^{cf}(x)>0\) on the relevant support, the right-hand side is strictly increasing in \(x^{cf, out}\), so the endpoint is unique.}   
\begin{equation}
    \int_{x^*+\epsilon}^{x^{kp,out}}
    h^{kp}(z)\,dz
   =
    \int_{x^*+\Delta x^*+\epsilon}^{x^{cf,out}}
    h^{cf}(x)\,dx,   
 \label{eq:xcf-shift-end}   
\end{equation}
For implementation, the same endpoint can be obtained from the all-agent mass balance:\footnote{\cite{chetty2011adjustment} used the equal-mass condition to infer how much we should move up the observed density to obtain the counterfactual density, so that the total observations under the observed and the counterfactual distributions are the same.}
\begin{equation}
  \int_{x^{in}}^{x^{kp,out}}    h^{kp}(z)\,dz    =   \int_{x^{in}}^{x^{cf,out}}   h^{cf}(x)\,dx. \label{eq:xcf-all-end}
\end{equation}
Equation \eqref{eq:xcf-all-end} is algebraically equivalent to equation \eqref{eq:xcf-shift-end} becuase assignments below $x^*$ are unaffected and equation \eqref{eq:B-M} accounts for the mass of bunchers that collapses to the cutoff.

\begin{figure}[!ht]
\centering
\caption{Changes in the Density Distribution}
\hypertarget{fig_density}{}
\includegraphics[width=0.75\linewidth]{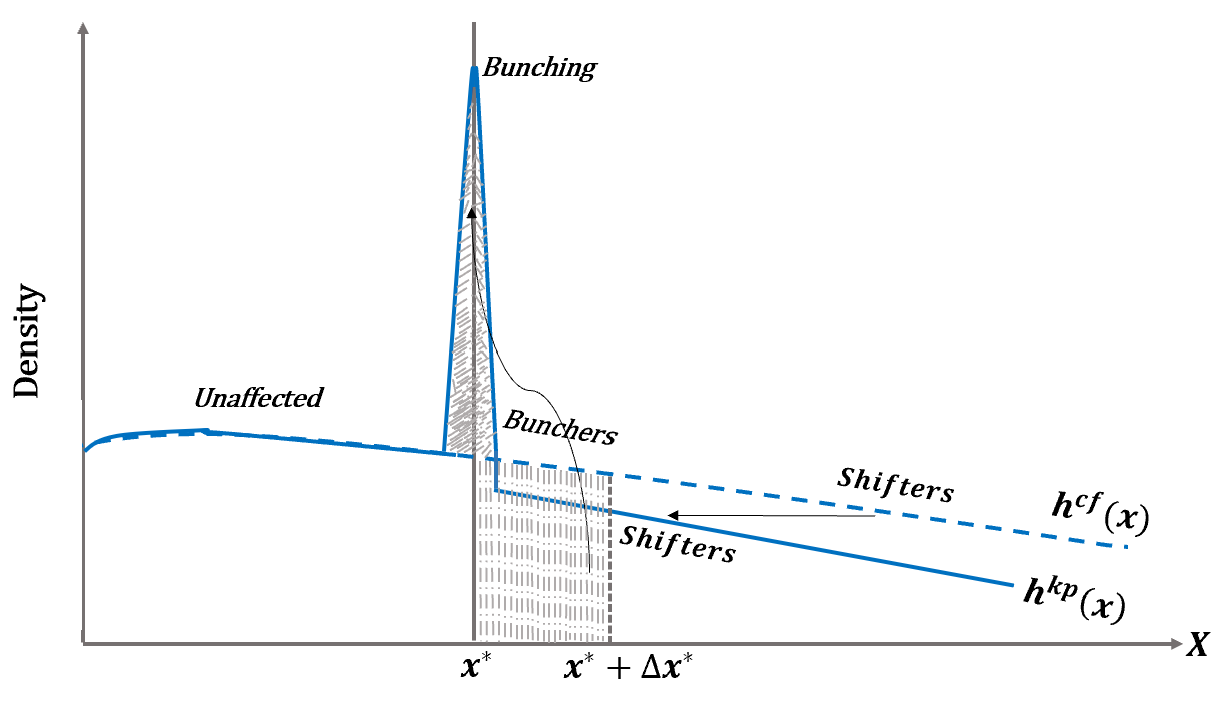}
\begin{minipage}{16cm}
Note: The figure illustrates the density of the assignment variable $X$. The solid line is the observed density under the kinked policy, $h^{kp}(x)$, with excess bunching at the threshold $x^*$. The dashed line is the counterfactual density $h^{cf}(x)$: it coincides with $h^{kp}(x)$ in the unaffected region below $x^*$; its segment over $(x^*,x^*+\Delta x^*]$ gives the counterfactual locations of \textit{bunchers}, who collapse to $x^*$; and its segment above $x^*+\Delta x^*$ is the counterfactual density of \textit{shifters}.
\end{minipage}
\label{fig:density-schematic}
\end{figure}

Figure~\ref{fig:density-schematic} illustrates the assignment-side objects: the observed density $h^{kp}$ with its excess bunching at $x^*$, and the counterfactual density $h^{cf}$ split into the buncher interval $(x^*,x^*+\Delta x^*]$ and the shifter region above it.


\subsection{Potential Outcomes and Total Effects}
\label{sec:potential-outcomes}

Let \(Y\) denote an outcome other than the assignment variable. The outcome may respond to the kinked policy through two channels: first, the kink changes the agent's assignment value from \(X^{cf}\) to \(X^{kp}\); second, the kink changes the payment associated with the assignment value.

For an agent of type \(\eta\), define the potential outcome under the counterfactual linear policy as
\[
Y^{cf}(\eta)
\equiv
Y\!\left(
X^{cf}(\eta),
T^{cf}(X^{cf}(\eta)),
\eta
\right),
\]
and the potential outcome under the kinked policy as
\[
Y^{kp}(\eta)
\equiv
Y\!\left(
X^{kp}(\eta),
T^{kp}(X^{kp}(\eta)),
\eta
\right).
\]
The main target of the paper is this total effect on outcomes of interest. It includes both the effect operating through the change in the assignment variable \(X\) and the effect operating through the change in the payment schedule \(T\).\footnote{A decomposition into assignment-variable and transfer channels can be useful for interpretation and welfare analysis, but it is not required for defining or identifying the aggregate total effects below.}

For unaffected agents with \(X^{cf}<x^*\), the kinked and counterfactual policies coincide locally:
\[
    X^{kp}=X^{cf},
    \quad
    T^{kp}(X^{kp})=T^{cf}(X^{cf}),    \qquad \Rightarrow  Y^{kp}=Y^{cf},
    \quad X^{cf}<x^*.
\]

For bunchers and shifters, the kink changes the assignment value, the payment schedule, or both. Their counterfactual outcomes are not directly observed.

\subsection{Counterfactual Outcome Objects}
\label{sec:outcome-functions}

Let 
\[
m^{cf}(x)
\equiv
\mathbb E\!\left[Y^{cf}\mid X^{cf}=x\right].
\]
denote the unobserved conditional outcome mean under the counterfactual policy.
This function describes the outcome that would have been observed for agents with latent counterfactual assignment value \(x\) under the linear policy. 

Let  \[
    m^{kp}(z)
    \equiv
    \mathbb E\!\left[Y^{kp}\mid X^{kp}=z\right]
\]
denote the conditional outcome mean mean under the kinked policy. This function describes  the observed outcome for agents with assignment value \(z\) under the kinked policy.

\begin{figure}[!ht]
\centering
\caption{Changes in the Conditional Outcome Mean Function}
\hypertarget{fig_y}{}
\includegraphics[width=0.75\linewidth]{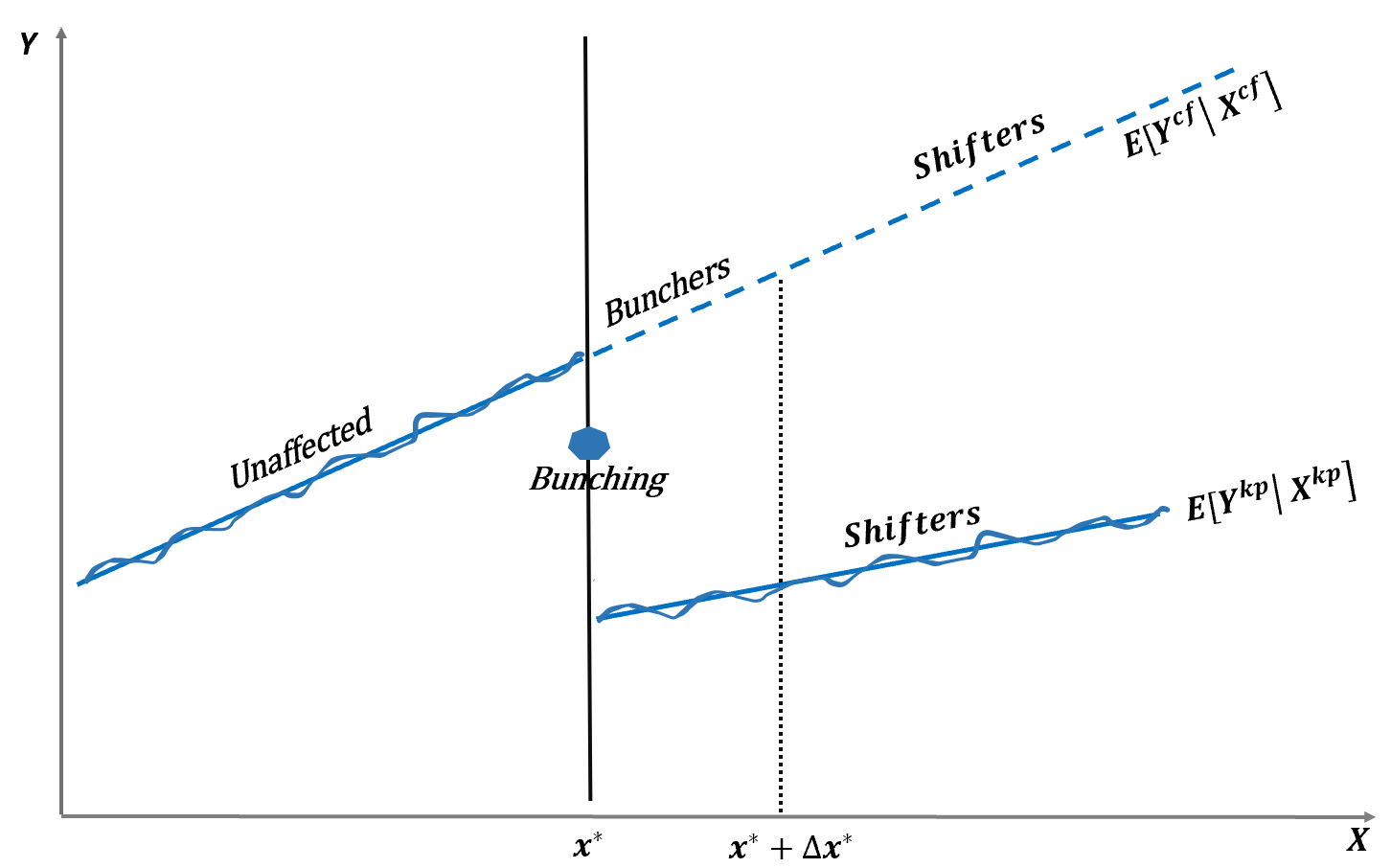}
\begin{minipage}{16cm}
Note: The figure illustrates the conditional outcome mean as a function of the assignment variable. The solid line is the observed outcome function under the kinked policy, $m^{kp}(x)=\mathbb E[Y^{kp}\mid X^{kp}=x]$; the dashed line is the counterfactual outcome function under the linear policy, $m^{cf}(x)=\mathbb E[Y^{cf}\mid X^{cf}=x]$. In the unaffected region below $x^*$ the two coincide. Bunchers collapse to the threshold $x^*$ (the marked point); their counterfactual outcomes lie on the dashed line over $(x^*,x^*+\Delta x^*]$. Shifters are observed above $x^*$ on the lower solid branch but originate from counterfactual values above $x^*+\Delta x^*$ on the dashed line. 
\end{minipage}
\label{fig:outcome-schematic}
\end{figure}

Figure~\ref{fig:outcome-schematic} illustrates these two functions and the three behavioral groups.

\subsection{Treatment-Effect Estimands}
\label{sec:estimands}

The framework distinguishes average treatment effects for bunchers and shifters. 

\paragraph{Bunchers.} Bunching agents come from the region with $X^{cf}\in (x^*, x^*+\Delta x^*]$ and bunch at the cutoff with $X^{kp}=x^*$ under the kinked policy. 
The counterfactual mean outcome for bunchers is
\begin{equation}
    \bar Y_B^{cf}
    =
    \frac{
    \int_{x^*}^{x^*+\Delta x^*}
    m^{cf}(x)h^{cf}(x)\,dx
    }{
    \int_{x^*}^{x^*+\Delta x^*}
    h^{cf}(x)\,dx
    }.
    \label{eq:YB-cf-section2}
\end{equation}
The treated mean outcome for bunchers is recovered from changes at the cutoff:
\begin{equation}
    \bar Y_B^{kp}
    =
    \frac{  m^{kp}(x^*) h^{kp}(x^*) -  m^{cf}(x^*)h^{cf}(x^*) }{  h^{kp}(x^*) - h^{cf}(x^*)}.
    \label{eq:YB-kp-section2}
\end{equation}

 The average treatment effect on bunchers is
\begin{equation}
    \tau_B
    =
    \bar Y_B^{kp}-\bar Y_B^{cf}.
    \label{eq:tau-B-section2}
\end{equation}

\paragraph{Shifters.} Agents with $X^{cf}>x^*+\Delta x^*$ reduce their assignment value but stay above the cutoff with $X^{kp}>x^*$.  
To define the treatment effect, we fix an observed evaluation endpoint \( x^{kp,out}>x^*\), such as the upper endpoint of the empirical plot window.

The treated mean outcome for shifters in the observed window $X^{kp} \in (x^*,x^{kp,out}]$ is
\begin{equation}
    \bar Y_S^{kp}
    =
    \frac{
    \int_{x^*}^{x^{kp,out}}
    m^{kp}(z)h^{kp}(z)\,dz
    }{
    \int_{x^*}^{x^{kp,out}}
    h^{kp}(z)\,dz
    }.
    \label{eq:YS-kp-section2}
\end{equation}
For the same shifter population, their counterfactual assingment range is $X^{cf} \in (x^*+\Delta x^*, x^{cf,out}]$ from equation (\ref{eq:xcf-shift-end}), and hence the counterfactual mean outcome is
\begin{equation}
    \bar Y_S^{cf}
    =
    \frac{
    \int_{x^*+\Delta x^*}^{x^{cf,out}}
    m^{cf}(x)h^{cf}(x)\,dx
    }{
    \int_{x^*+\Delta x^*}^{x^{cf,out}}
    h^{cf}(x)\,dx
    }.
    \label{eq:YS-cf-section2}
\end{equation}
The average treatment effect on shifters in the evaluation window is
\begin{equation}
    \tau_S
    =
    \bar Y_S^{kp}
    -
    \bar Y_S^{cf}.
    \label{eq:tau-S-section2}
\end{equation}
This estimand compares average treated and counterfactual outcomes for the same set of shifters. The observed set is defined by \(X^{kp}\in(x^*,x^{kp,out}]\). The counterfactual set is \(X^{cf}\in(x^*+\Delta x^*,x^{cf, out}]\). Monotonicity links these two sets; the equal-mass condition identifies the counterfactual endpoint.

\subsection{What Remains to Be Identified}
\label{sec:identification-overview}

The definitions above clarify what must be recovered from the data. First, one must recover the  counterfactual assignment density \(h^{cf}(x)\) and the marginal bunching response \(\Delta x^*\). If \(h^{cf}(x)\) were known near the kink, the buncher mass-balance condition in \eqref{eq:B-M} would identify \(\Delta x^*\). 
For any observed endpoint \(x^{kp,out}\) of shifters, the equal-mass condition in \eqref{eq:xcf-shift-end}-\eqref{eq:xcf-all-end} then identifies the counterfactual endpoint \(x^{cf,out}\).   
Second, one must recover the counterfactual conditional outcome mean function \(m_F^{cf}(x)\) over the missing buncher interval and the shifter counterfactual interval. 
Section~\ref{sec:assignment-identification} addresses the assignment-side problem, and Section~\ref{sec:outcome-identification} addresses the outcome-side problem.

\section{Identification of the Counterfactual Assignment Distribution}
\label{sec:assignment-identification}

Section~\ref{sec:framework} defined the latent counterfactual assignment value $X^{cf}$, the observed assignment value under the kinked policy $X^{kp}$, the marginal bunching response $\Delta x^*$, and the three behavioral groups induced by the kink. This section identifies the assignment-side objects needed for the outcome analysis: the focal group's counterfactual assignment density $h_F^{cf}(x)$ and the marginal bunching response $\Delta x^*$. 
The outcome-side problem is left to Section~\ref{sec:outcome-identification}. The goal here is only to recover where affected agents would have located under the counterfactual linear policy.

\subsection{Bunching mass and assignment-side non-identification}
\label{sec:bunching-mass-nonid}

Under the behavioral mapping in Section~\ref{sec:potential-assignments}, agents with counterfactual assignment values in the interval $(x^*,x^*+\Delta x^*]$ bunch at the kink. The excess bunching mass satisfies $B= \int_{x^*}^{x^*+\Delta x^*} h^{cf}(x)\,dx .$ 
Here $B$ is the excess mass at the kink. 
With a single kink and an unrestricted counterfactual density,  $h_F^{cf}(x)$ and $\Delta x^*$ are not separately identified \citep{blomquist2021bunching}. 
Identification therefore requires  additional information about the shape of the counterfactual assignment distribution \citep{bertanha2023better, goff2022treatment}.


\subsection{Design-Assisted Density Identification}
\label{sec:density-shape-restriction}  \label{sec:density-identification}

Our main specification uses policy variations to discipline the local shape of the focal group's counterfactual density. Let \(F\) denote the focal group that is subject to the kink analysis above.  Let \(P\) denote a placebo group that is \textit{not} subject to the focal kink in a local window \(\mathcal W\) around \(x^*\). The placebo group may come from another population or another time period whose relevant threshold lies outside \(\mathcal W\).
Let \(h_F^{kp}(x), h_F^{cf}(x) \)  denote the observed/counterfactual densities of the assignment variable for the focal group, and \(h_P(x)\) denote the observed density for the placebo group.

\begin{assumption}[Design-assisted density shape]
\label{assume:density-shape}
Let $\mathcal W$ be a local window around $x^*$, and let $q_h(x)$ be a known vector of low-dimensional basis functions. In $\mathcal W$, the focal group's counterfactual density and the placebo density satisfy
\begin{equation}
    h_F^{cf}(x)-h_P(x)=q_h(x)'\delta_h,
    \qquad x\in\mathcal W.
    \label{eq:density-shape}
\end{equation}
In the baseline specification, $q_h(x)=(1,x)'$, so the focal and placebo groups may differ in both density level and local trend. 
\end{assumption}

\begin{figure}[!ht]
\centering
\caption{Design-based counterfactual density.}
\includegraphics[width=0.7\linewidth]{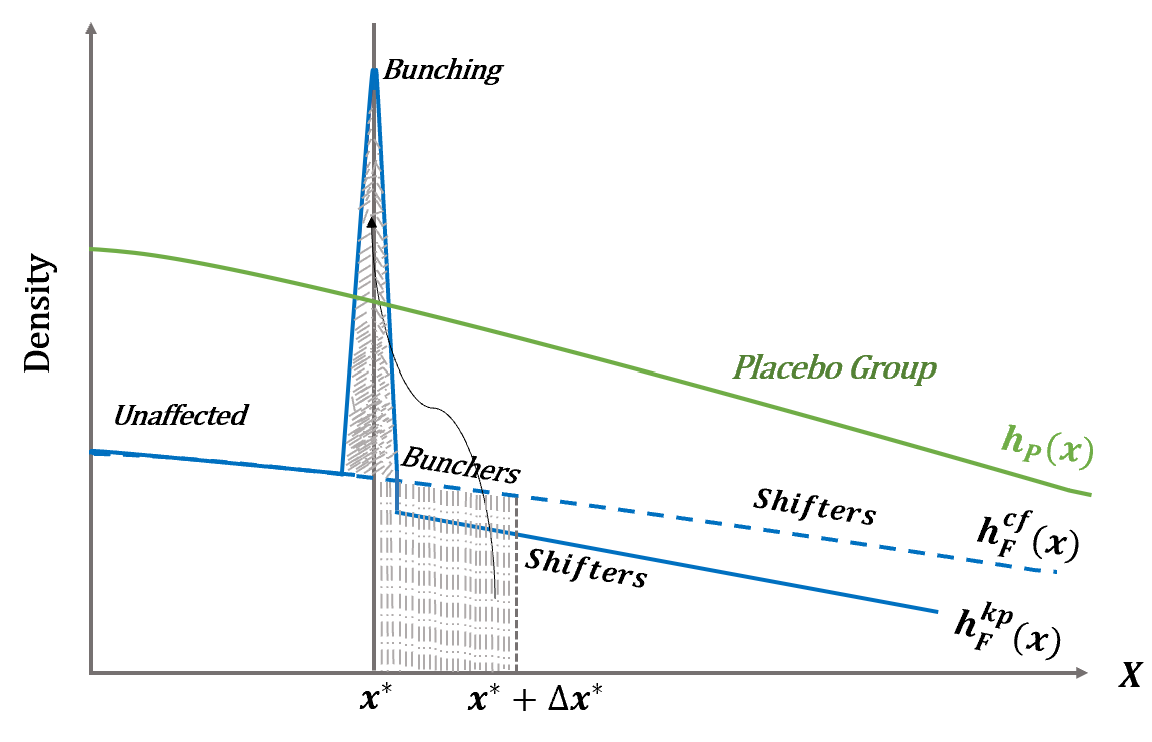}
\begin{minipage}{16cm}
\footnotesize{Note: The blue curves are the focal group ($F$): the observed density $h^{kp}_F(x)$ (solid) with excess bunching at $x^*$, and the counterfactual density $h^{cf}_F(x)$ (dashed). The green curve is the density for placebo group ($P$), which is \textit{not} subject to the focal kink and is smooth through $x^*$. The placebo may differ from the focal group in level and slope; Assumption~\ref{assume:density-shape} requires only that, after a level-and-slope adjustment estimated on the unaffected region  $X<x^*$, 
the focal counterfactual density  share the same local shape as the placebo group---which identifies $h^{cf}_F(x)$ above $x^*$.}  
\end{minipage}
\label{fig:density-design}
\end{figure}

Assumption~\ref{assume:density-shape} is a local parallel-shape condition. With \(q_h(x)=(1,x)'\), the substantive restriction is that, after allowing the focal counterfactual and placebo densities to differ in level and slope, there is no unobserved bump, discontinuity, or curvature change in the missing interval that is present for the focal group but absent from the placebo group.
Figure~\ref{fig:density-design} illustrates the restriction. 

$\mathcal U=\mathcal W\cap\{x<x^*\}$ denotes the unaffected region. For $x\in\mathcal U$, agents face the same marginal incentive under the kinked and counterfactual policies; thus,
\(
    X^{kp}=X^{cf},
    h_F^{kp}(x)=h_F^{cf}(x),
     \forall x\in\mathcal U.
    \label{eq:unaffected-density-equality}
\)
Combined with Assumption~\ref{assume:density-shape} gives
\begin{equation}
    h_F^{kp}(x)-h_P(x)=q_h(x)'\delta_h,
    \quad x\in\mathcal U.
    \label{eq:delta-h-identification}
\end{equation}
Thus, $\widehat{\delta}_h$ is identified from the focal group's own unaffected observations and the placebo density.
Given $\widehat{\delta}_h$, the focal group's counterfactual density is identified throughout the local window by:
\begin{equation}
    \widehat{h}_F^{cf}(x)  = h_P(x)+q_h(x)'\widehat{\delta}_h,
    \qquad x\in\mathcal W.
    \label{eq:hcf-identified}
\end{equation}

The marginal bunching response is then pinned down by  mass balance. Let $B_F$ denote the focal group's excess bunching mass. Define the missing-mass function
\begin{equation*}
    M(\Delta)
    =
    \int_{x^*}^{x^*+\Delta}
    \left[h_P(x)+q_h(x)'\widehat{\delta}_h\right]dx .
    \label{eq:missing-mass-function}
\end{equation*}
The true marginal bunching response \(\Delta x^*\) solves
\begin{equation}
    B_F=M(\Delta x^*).
    \label{eq:delta-mass-balance}
\end{equation}

\begin{assumption}[Density support, rank, and positivity]
\label{assume:density-regularity} 
Let $[0,\bar\Delta]$ be the admissible set for the marginal bunching response $\Delta x^*$, with $[x^*,x^*+\bar\Delta]\subset\mathcal W$. The following conditions hold.
\begin{enumerate}
    \item The placebo density $h_P(x)$ is observed, integrable, and continuous on $\mathcal W$.
    \item The unaffected region $\mathcal U$ has positive length, and $q_h(x)$ has full rank on $\mathcal U$: for some positive weight function $w(x)$,
   $\int_{\mathcal U} q_h(x)q_h(x)'w(x)\,dx$ is nonsingular.
    \item The implied focal counterfactual density is positive on the relevant missing region:   $ h_P(x)+q_h(x)'\delta_h>0,     \quad x\in[x^*,x^*+\bar\Delta].$  
  \item The observed bunching mass lies in the feasible range: 
    $        0\leq B_F\leq     \int_{x^*}^{x^*+\bar\Delta}   \left[h_P(x)+q_h(x)'\delta_h\right]dx .  $
\end{enumerate}
\end{assumption}
\noindent 
These regularity conditions ensure that the adjustment parameter is identified, that the implied counterfactual density is a valid local density, and that the mass-balance equation has a solution in the admissible range.  The substantive identifying content is Assumption~\ref{assume:density-shape}.


\begin{proposition}[Assignment-side identification]
\label{prop:assignment-identification}
Suppose the behavioral grouping in Section~\ref{sec:potential-assignments} and Assumptions~\ref{assume:density-shape}--\ref{assume:density-regularity} hold. Then:
\begin{enumerate}
    \item The adjustment parameter \(\delta_h\) is identified from the unaffected region \(\mathcal U\).
    \item The focal group's counterfactual density \(h_F^{cf}(x)\) is identified on \(\mathcal W\) by \eqref{eq:delta-h-identification} \&  \eqref{eq:hcf-identified}. 
    \item The missing-mass function \(M(\Delta)\) is continuous and strictly increasing on \(\mathcal D\).
    \item The mass-balance equation \(B_F=M(\Delta x^*)\) has a unique solution in \(\mathcal D\), which identifies \(\Delta x^*\).
\end{enumerate}
\end{proposition}

\textit{\textbf{Intuition.}} The unaffected region reveals how the focal and placebo densities differ when both groups are locally unaffected by the focal kink. Assumption~\ref{assume:density-shape} extends this low-dimensional difference into the missing region above the kink. Once the counterfactual density in that region is known, the observed excess bunching mass pins down the interval that have collapsed to the kink.

\subsection{Assignment-side diagnostics}
\label{sec:assignment-diagnostics}

The identifying restriction in Assumption~\ref{assume:density-shape} has empirical implications. First, in the unaffected region, the residualized density difference
\(
    h_F^{kp}(x)-h_P(x)-q_h(x)'\delta_h
\)
should not display systematic curvature, jumps, or local kinks. 
Second, if placebo thresholds are available, the same procedure can be applied at values of \(x\) that are not thresholds for the focal group. At those placebo thresholds, the method should not find excess bunching or a missing region.  
Third, when multiple placebo groups or policy years are available, check robustness using alternative placebo groups. 

\begin{remark}
The placebo density in Assumption~\ref{assume:density-shape} can be replaced by a predetermined-covariate reweighted placebo density \(h_P^\omega(x)\) that matches the focal group's covariate distribution in the unaffected region \(\mathcal U\) The estriction becomes \(h_F^{cf}(x)-h_P^\omega(x)=q_h(x)'\delta_h\). This is analogous to matching before difference-in-differences: reweighting  makes the condition more credible.  
\end{remark}

\section{Identification of the Counterfactual Outcome Function and Treatment Effects}
\label{sec:outcome-identification}

Section~\ref{sec:assignment-identification} identifies the assignment-side objects required for causal inference: the focal group's counterfactual assignment density \(h_F^{cf}(x)\) and the marginal bunching response \(\Delta x^*\).
This section identifies the outcome-side objects. The central object is the counterfactual outcome function
\(
    m_F^{cf}(x)=\mathbb E[Y^{cf}\mid X^{cf}=x,F].
\)


\subsection{Outcome-side non-identification after relocation}
\label{sec:outcome-nonid}
Section  ~\ref{sec:assignment-identification} which counterfactual assignment values $X^{cf}$ correspond to bunching and shifting agents, respectively. However, they do not reveal what outcomes $Y^{cf}$ those agents would have had under the counterfactual policy.

This distinction remains even under a stronger pointwise relocation rule. Suppose, for illustration, that a known monotone map $X^{kp}=R(X^{cf})$ allowed each observed shifter to be indexed by her latent cpunterfactual value.  The treated data would then reveal the auxiliary conditional outcome mean function $m^r_F(x)\equiv E[Y^{kp}|X^{cf}=x, F], \quad x>x^*+\Delta x^*$. 
 But in the shifter region,
 \[m^r_F(x)=m^{cf}_F(x)+\tau(x).\] 
If both the counterfactual conditional mean function $m^{cf}_F(x)$ and the treatment effect $\tau(x)$ are unrestricted,  many pairs rationalize the same $m^r_F(x)$, leading to non-identification issue. 

The unaffected region provides the first source of information. Recall that 
$\mathcal U=\mathcal W\cap\{x<x^*\}$
denote the part of the local window in which focal-group agents face the same policy under the kinked and counterfactual linear schedules. For these agents, $  X^{kp}=X^{cf},  \quad  Y^{kp}=Y^{cf}.$ 
Hence the observed focal outcome function identifies the counterfactual outcome function below the kink:
\begin{equation}
    m^{kp}_F(x)=m^{cf}_F(x),
    \qquad x\in\mathcal U,
    \label{eq:unaffected-outcome-equality}
\end{equation}
where \(m^{kp}_F(x)=\mathbb E[Y^{kp}\mid X^{kp}=x, F]\). The identification problem is how to extend this information from the unaffected region into the missing counterfactual region above the kink.

\subsection{Design-assisted outcome shape}
\label{sec:outcome-shape-restriction} \label{sec:mcf-identification}
Similarly, we use policy variations to discipline the local shape of the focal group's counterfactual outcome function. 
Let \(m_F^{cf}(x)=\mathbb E[Y^{cf}|X^{cf}=x, F]\) denotes the counterfactual conditional outcome mean for the focal group $F$, and \(m_F^{kp}(z)=\mathbb E[Y^{kp}|X^{kp}=z]\) denotes the observed conditional mean under the kinked policy. 
Let \(    m_P(x)=\mathbb E[Y\mid X=x,P]\) denote the observed conditional outcome mean for placebo group $P$ that is not subject to the focal kink in the local window \(\mathcal W\).


\begin{assumption}[Design-assisted Outcome Identification]
\label{assume:outcome-shape}  

Let \(\mathcal W\) be a local window around \(x^*\), and let \(q_y(x)\) be a known vector of low-dimensional basis functions. In \(\mathcal W\), the focal group's counterfactual outcome function and the placebo group's observed outcome function satisfy
\begin{equation}
    m_F^{cf}(x)-m_P(x)=q_y(x)'\delta_y,
    \qquad x\in\mathcal W.
    \label{eq:outcome-shape}
\end{equation}
In the baseline specification, \(q_y(x)=(1,x)'\), so the focal and comparison groups may differ in both outcome level and local outcome trend.
\end{assumption}


Assumption~\ref{assume:outcome-shape} is a local parallel-shape condition for outcomes. It does not require the placebo group to have the same outcome level or slope as the focal group.  With \(q_y(x)=(1,x)'\), the placebo group disciplines the remaining local curvature of the counterfactual outcome function.  The substantive restriction is that, after allowing the focal and placebo groups to differ in outcome level and local outcome trend, there is no unobserved bend, jump, or curvature change in the focal group's counterfactual outcome function that is absent from the placebo group.

Combining the unaffected-region equality in \eqref{eq:unaffected-outcome-equality} with Assumption~\ref{assume:outcome-shape} gives
\begin{equation}
    m_F^{kp}(x)-m_P(x)=q_y(x)'\delta_y,
    \qquad x\in\mathcal U.
    \label{eq:delta-y-identification}
\end{equation}
Thus, the focal group's own unaffected observations identify the adjustment \(\widehat{\delta}_y\). Once \(\widehat{\delta}_y\) is known, the counterfactual outcome function is identified throughout the local window:
\begin{equation}
    m_F^{cf}(x)
    =
    m_P(x)+q_y(x)'\delta_y,
    \qquad x\in\mathcal W.
    \label{eq:mcf-identified}
\end{equation}

\begin{figure}[!ht]
\centering
\caption{Design-based counterfactual outcome function.}
\includegraphics[width=0.7\linewidth]{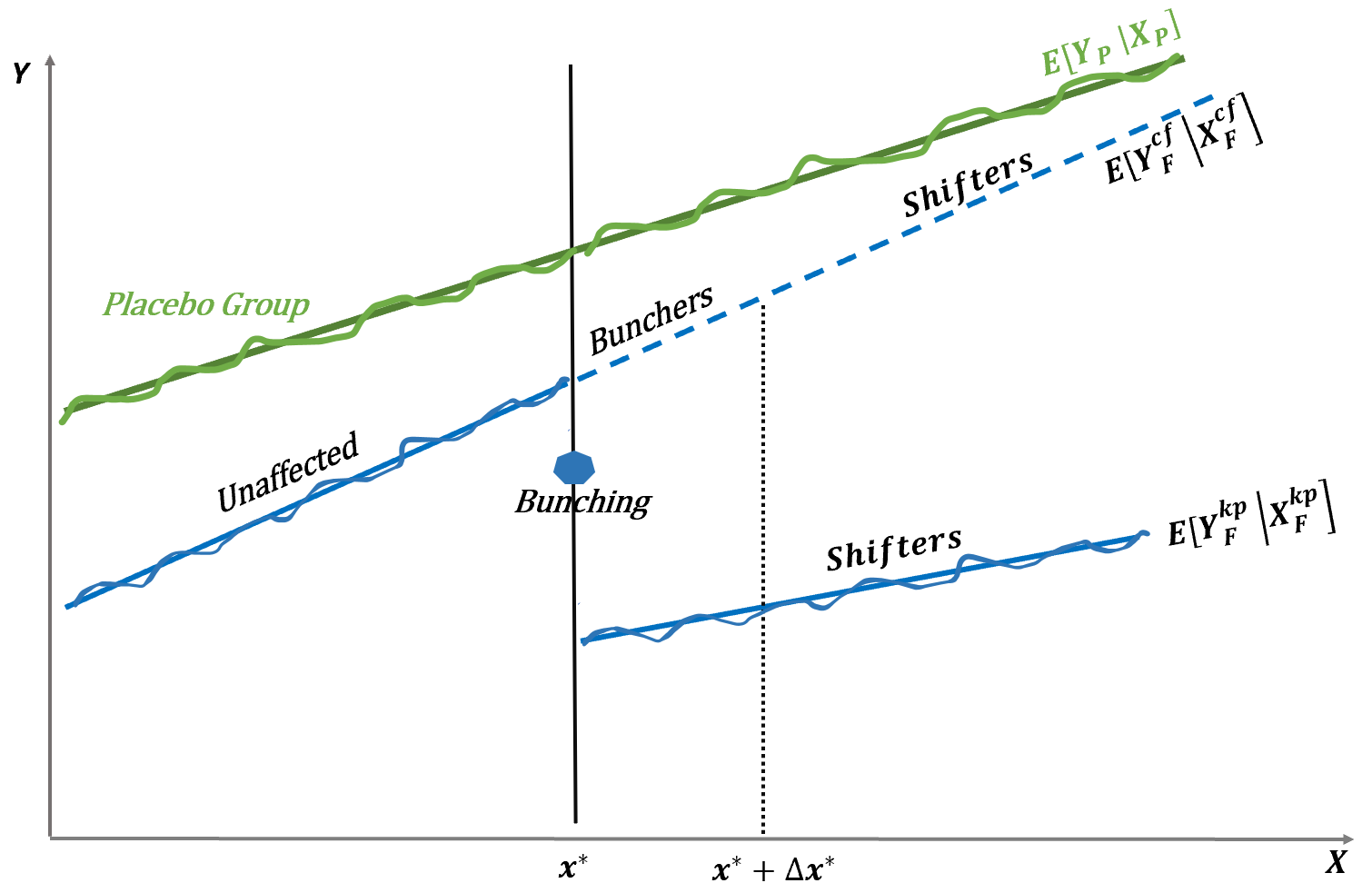}
\begin{minipage}{16cm}
\footnotesize{Note: The blue curves are the focal group: the observed outcome function $m_F^{kp}(x)=\mathbb E[Y_F^{kp}\mid X_F^{kp}=x]$ (solid), with bunchers collapsed to $x^*$ and shifters on the lower branch above $x^*$, and the counterfactual outcome function $m_F^{cf}(x)=\mathbb E[Y_F^{cf}\mid X_F^{cf}=x]$ (dashed). The green curve is the placebo group's observed outcome function $m_P(x)=\mathbb E[Y_P\mid X_P=x]$, which is \textit{not} subject to the focal kink. The placebo may differ from the focal group in level and slope; Assumption~\ref{assume:outcome-shape} requires only that, after a level-and-slope adjustment estimated on the unaffected region $X<x^*$, the focal counterfactual outcome function share the same local shape as the placebo group---which identifies $m_F^{cf}(x)$ above $x^*$.} 
\end{minipage}
\label{fig:outcome-design}
\end{figure}

\begin{assumption}[Outcome support and rank]
\label{assume:outcome-regularity}
The following conditions hold in local window \(\mathcal W\):
\begin{enumerate}
    \item The placebo's outcome function \(m_P(x)\) is observed and continuous on \(\mathcal W\).
    \item The unaffected region \(\mathcal U\) has positive length, and \(q_y(x)\) has full rank on \(\mathcal U\). That is, for some positive weight function \(\omega_y(x)\),
   $ \int_{\mathcal U} q_y(x)q_y(x)'\omega_y(x)\,dx$ is nonsingular.
\end{enumerate}
\end{assumption}
\noindent 
The substantive identifying content is Assumption~\ref{assume:outcome-shape}. Assumption~\ref{assume:outcome-regularity} ensures that the low-dimensional adjustment is identified from the unaffected region.   
Figure~\ref{fig:outcome-design} illustrates this restriction: the placebo group disciplines the shape of the focal group's counterfactual outcome function $m_F^{cf}$. 

\begin{proposition}[Outcome-side identification]
\label{prop:outcome-identification}  
Suppose the assignment-side objects \(h_F^{cf}(x)\) and \(\Delta x^*\) are identified as in Section~\ref{sec:assignment-identification}. Suppose Assumptions~\ref{assume:outcome-shape}--\ref{assume:outcome-regularity} hold. Then:
\begin{enumerate}
    \item The adjustment parameter \(\delta_y\) is identified from the unaffected region \(\mathcal U\).
    \item The counterfactual outcome function \(m_F^{cf}(x)\) is identified on \(\mathcal W\) by equations \eqref{eq:delta-y-identification}-\eqref{eq:mcf-identified}.
\end{enumerate}
\end{proposition}


\subsection{Identification of treatment effects}
\label{sec:treatment-effect-identification}

\textbf{Bunchers.} 
Their counterfactual mean outcome is identified by integrating the identified counterfactual outcome function over the missing interval:
\begin{equation}
    \bar Y_B^{cf}
    =
    \frac{
    \int_{x^*}^{x^*+\Delta x^*}
    m_F^{cf}(x)h_F^{cf}(x)\,dx
    }{
    \int_{x^*}^{x^*+\Delta x^*}
    h_F^{cf}(x)\,dx
    }.
    \label{eq:buncher-ycf-identified}
\end{equation}

Their treated mean outcome is associated with the excess mass at the kink.

In the \textit{sharp-bunching} benchmark, all bunchers locate at \(x^*\).  The excess mass at $x^*$ is generated by \textit{bunchers}, i.e., $B_F=h_F^{kp}(x^*)-h_F^{cf}(x^*)$.  
The treated mean outcome for\textit{ bunchers} is 
\begin{equation}\label{eq:sharp-ykp}
   \bar Y_B^{kp}
    =\frac{m_F^{kp}(x^*)h^{kp}_F(x^*)-m_F^{cf}(x^*)h_F^{cf}(x^*)}{h_F^{kp}(x^*)-h_F^{cf}(x^*)},
\end{equation}

In  \textit{diffuse-bunching} implementations, agents may locate  around the kink $[x^*-u_1,x^*+u_2].$ 
The observed density in this region may contain both bunchers and non-bunchers. We therefore recover the treated mean for bunchers by subtracting the non-buncher component from the observed. 
   \begin{itemize}
       \item  For $x\in [x^*-u_1,x^*]$, the observed consist of \textit{unaffected} agents and \textit{bunchers}. The density of bunchers at observed assignment value $x$ is $h_B^{kp}(x) =h_F^{kp}(x)-h_F^{cf}(x),$ and their treated mean outcome is 
\begin{equation}\label{eq:diffuse-ykp-l}
       m_B^{kp}(x)=\frac{m_F^{kp}(x)h_F^{kp}(x)-m_F^{cf}(x)h_F^{cf}(x)}{h_F^{kp}(x)-h_F^{cf}(x)}.   
      \end{equation}

    \item 
For $x\in (x^*,x^*+u_2]$, the observed consists of \textit{shifters} and \textit{bunchers}. Let $h_S^{kp}(x)$ and $m_S^{kp}(x)$ denote the density and treated mean outcome of shifters at observed assignment value $x$.\footnote{Shifters are directly observed in the region with $X^{kp}>x^*+u_2$ and we can extrapolate it to the region $(x^*, x^*+u_2]$ given that $u_2$ is generally small.}
The density of bunchers in this segment is
$h_B^{kp}(x) =h_F^{kp}(x)-h_S^{kp}(x),$
and their treated mean outcome is
\begin{equation}\label{eq:diffuse-ykp-r}
m_B^{kp}(x)=\frac{m_F^{kp}(x)h_F^{kp}(x)-m_S^{kp}(x)h_S^{kp}(x)}{h_F^{kp}(x)-h_S^{kp}(x)}.    
\end{equation}
\item Combining the left and right segments, the average treated outcome for bunchers under diffuse bunching is
\begin{equation}  \label{eq:diffuse-ykp}
 \bar Y_B^{kp}
 \equiv   E\left[Y^{kp}\mid B,F\right]
=\frac{\int_{x^*-u_1}^{x^*+u_2}m_B^{kp}(x)h_B^{kp}(x)\,dx}{\int_{x^*-u_1}^{x^*+u_2}h_B^{kp}(x)\,dx}.  
\end{equation}
where excess mass $B_F=\sum_{x^*-u_1}^{x^*} \big[ h^{kp}_F(x)-h^{cf}_F(x)\big] +\sum_{x^*+\epsilon}^{x^*+u_2} \big[h^{kp}_F(x)-h_S(x)\big]$. 
 \end{itemize}


The average treatment effect on bunchers is therefore
\begin{equation}
    \tau_B
    =
    \bar Y_B^{kp}-\bar Y_B^{cf}.
    \label{eq:tau-b-identified}
\end{equation}

\textbf{Shifters.} 
Under \textit{sharp bunching}, shifters with $X^{cf}>X^*+\Delta x^*$ reduce their assignment value but stay above the cutoff with $X^{kp}>x^*$.
The average effect on shifters can then be estimated:
\begin{equation}
    \tau_S
    =\frac{ \int_{x^*+\epsilon}^{x^{kp,out}}m_F^{kp}(x) h_F^{kp}(x)\,dx }{ \int_{x^*+\epsilon}^{x^{kp,out}} h_F^{kp}(x)\,dx } -
\frac{ \int_{x^*+\Delta x^*+\epsilon}^{x^{cf, out}}m_F^{cf}(x) h_F^{cf}(x)\,dx }{ \int_{x^*+\Delta x^*}^{x^{cf, out}} h_F^{cf}(x)\,dx }
    \label{eq:tau-s-identified}
\end{equation}
where $\int_{x^*+\epsilon}^{x^{kp,out}} h_F^{kp}(x)\,dx =\int_{x^*+\Delta x^*}^{x^{cf,out}} h_F^{cf}(x)\,dx$ denotes the number of shifters in the evaluation window, with $x^{kp,out}$ pre-chosen and $x^{cf,out}$ computed from the mass balance equation \eqref{eq:xcf-shift-end} or \eqref{eq:xcf-all-end}.

Under \textit{diffuse bunching}, the region $X^{kp}\in (X^*, X^*+u_2]$ contains both bunching and shifting agents. The region $X^{kp}>X^*+u_2$ only contains shifters, that is,  $h_S^{kp}(x)=h_F^{kp}(x), \quad x>x^*+u_2$. Hence, we can extrapolate it to the diffusion region $(x^*, x^*+u_2]$ and obtain $h_S^{kp}(x), \quad x>x^*$, given that $u_2$ is generally small. Similarly, we have $m_S^{kp}(x)\equiv \mathbb E[Y^{kp}|X^{kp}, S, F]=m^{kp}(x), \quad x>x^*+u_2$, and can extrapolate it to the diffusion region $(x^*, x^*+u_2]$ and obtain $h_S^{kp}(x), \quad x>x^*$.
 Then, we can estimate the average effect on shifters: 
 \begin{equation}
    \tau_S
    =\frac{ \int_{x^*+\epsilon}^{x^{kp,out}}m_S^{kp}(x) h_S^{kp}(x)\,dx }{ \int_{x^*+\epsilon}^{x^{kp,out}} h_S^{kp}(x)\,dx } -
\frac{ \int_{x^*+\Delta x^*+\epsilon}^{x^{cf, out}}m_F^{cf}(x) h_F^{cf}(x)\,dx }{ \int_{x^*+\Delta x^*}^{x^{cf, out}} h_F^{cf}(x)\,dx }
    \label{eq:tau-s-diffuse}
\end{equation}
where $\int_{x^*+\epsilon}^{x^{kp,out}} h_S^{kp}(x)\,dx =\int_{x^*+\Delta x^*}^{x^{cf,out}} h_F^{cf}(x)\,dx$.




\subsection{Outcome-side diagnostics}
\label{sec:outcome-diagnostics}

Assumption~\ref{assume:outcome-shape} has testable implications. First, in the unaffected region, the residualized outcome difference
$ m_F^{kp}(x)-m_P(x)-q_y(x)'\delta_y$
should not display systematic curvature, jumps, or local slope changes. 
Second, when placebo thresholds are available, the same outcome-shape restriction can be checked at locations where the focal group is not exposed to a kink. 
Third, we can run robustness checks by varying the placebo groups, when multiple placebo groups or policy years are available. The estimated counterfactual outcome function should be stable across placebo groups.

\begin{remark} 
If composition differences between the focal and placebo groups make the raw placebo outcome shape less comparable, one can replace \(m_P(x)\) with a reweighted placebo outcome function \(m_P^\omega(x)\) that matches the focal group's covariate distribution. The restriction becomes \(m_F^{cf}(x)-m_P^\omega(x)=q_y(x)'\delta_y\), strengthening the plausibility of the outcome shape comparison. 
\end{remark}





\section{Estimation and Inference}
\label{sec:estimation}

This section gives the sample analogue of Sections~\ref{sec:assignment-identification} and~\ref{sec:outcome-identification}. The goal is to make the procedure implementable. The researcher chooses a local window \(\mathcal W=[x^*-\ell,x^*+r]\), a bin width or smoothing bandwidth, a diffuse-bunching region \(\mathcal D=[x^*-u_1,x^*+u_2]\), an observed shifter endpoint \(x^{kp,out}\), a placebo group \(P\), and basis functions for the density and outcome restrictions, \(q_h(x)\) and \(q_y(x)\). All objects are then estimated within the same local design.



Let \(i=1,\ldots,n\) index observations. For the focal group, the observed assignment is \(X_i=X_i^{kp}\) and the observed outcome is \(Y_i=Y_i^{kp}\). For the placebo group, \(X_i\) and \(Y_i\) denote the corresponding assignment and outcome under a regime without the focal kink. Let \(\widehat h_F^{kp}(x)\), \(\widehat h_P(x)\), \(\widehat m_F^{kp}(x)\), and \(\widehat m_P(x)\) denote te binned or smoothed  observed density and outcome functions.


\subsection{Estimator}
\label{sec:baseline-estimator}

The estimator proceeds as follows.

\begin{enumerate}
\item \textit{Observed profiles.} Estimate the focal and placebo density profiles \(\widehat h_F^{kp}(x)\) and \(\widehat h_P(x)\) on \(\mathcal W\), and estimate the corresponding outcome profiles \(\widehat m_F^{kp}(x)\) and \(\widehat m_P(x)\). The same bins or smoothing rules are used for focal and placebo objects.


\item \textit{Assignment-side first stage.} 
 Fit the density adjustment on the unaffected region \(\mathcal U=\mathcal W\cap\{x<x^*\}\),
    \begin{equation}
        \widehat\delta_h
        =
        \arg\min_\delta
        \sum_{x_b\in\mathcal U}
        \left[
            \widehat h_F^{kp}(x_b)-\widehat h_P(x_b)-q_h(x_b)'\delta_h
        \right]^2,
        \label{eq:delta-h-estimator}
    \end{equation} 
where \(x_b\) indexes bins. The common shape $\widehat h_P(x_b)$ are approximated flexibly using OLS, possion, negative binomial specifications, etc. \(q_h(x)=(1,x)'\)  restrict \textit{only} the focal$-$placebo difference in level and slope.
The counterfactual density is
    \begin{equation}
        \widehat h_F^{cf}(x)
        =
        \widehat h_P(x)+q_h(x)'\widehat\delta_h,
        \qquad x\in\mathcal W.
        \label{eq:hcf-estimator}
    \end{equation}
 Estimate the excess bunching mass \(\widehat B_F\) using the  chosen bunching region $[x^*-u_1, x^*+u_2]$. Then, solve the sample analogue of the buncher mass-balance equation for the marginal bunching response \(\Delta x^*\):
    \begin{equation}
        \widehat B_F
        =
        \int_{x^*}^{x^*+\widehat{\Delta x}^{\,*}}
        \widehat h_F^{cf}(x)\,dx,
        \label{eq:delta-x-estimator}
    \end{equation}

For shifters, solve the equal-mass endpoint condition for the shifter's counterfactual endpoint $x^{cf,out}(x^{kp,out})$:
    \begin{equation}
        \int_{x^*}^{x^{kp,out}}\widehat h_{S}^{kp}(x)\,dx
        =
        \int_{x^*+\widehat{\Delta x}^{\,*}}^{\widehat x^{cf,out}}
        \widehat h_F^{cf}(x)\,dx.
        \label{eq:xcf-out-estimator}
    \end{equation}
Under sharp bunching, shifter's density under the kinked policy \(\widehat h_S^{kp}(x)=\widehat h_F^{kp}(x), \quad x>x^*\); under diffuse bunching, \(\widehat h_S^{kp}\) is the estimated non-buncher component in  the diffuse region $(x^*, x^*+u_2]$ and is $h_F^{kp}(x)$ when $x>x^*+u_2$.



 \item \textit{Outcome-side second stage and aggregation.} Fit the outcome adjustment on the same unaffected region,
    \begin{equation}
        \widehat \delta_y
        =
        \arg\min_{\delta_y}
        \sum_{x_b\in\mathcal U}      
        \left[
            \widehat m_F^{kp}(x_b)-\widehat m_P(x_b)-q_y(x_b)'\delta_y
        \right]^2,
        \label{eq:delta-y-estimator}  
    \end{equation}
where the common shape $\widehat m_P(x_b)$ can be appromiated using OLS, possion, negative binomial specifications, etc. \(q_y(x)=(1,x)'\)  restrict \textit{only} the focal$-$placebo difference in level and slope. 
   Construct
    \begin{equation}
        \widehat m_F^{cf}(x)
        =
        \widehat m_P(x)+q_y(x)'\widehat\delta_y,
        \qquad x\in\mathcal W.
        \label{eq:mcf-estimator}
    \end{equation}

The estimated treatment effects are plug-in versions of the estimands in Section~\ref{sec:estimands} and the formulas in Section~\ref{sec:outcome-identification}: 
    \begin{itemize}
    \item $\widehat\tau_B=\widehat{\bar Y}_B^{kp}-\widehat{\bar Y}_B^{cf}$ computes the average treatment effect for bunchers, where the average treated outcome for bunchers $\bar{Y}_B^{kp}$ is obtained via equations \eqref{eq:diffuse-ykp-l}--\eqref{eq:diffuse-ykp}, and their average counterfactual outcome $\bar{Y}_B^{cf}$ over the corresponding interval \(X^{cf}\in(x^*,x^*+\widehat{\Delta x}^{\,*}]\) is obtained via equation \eqref{eq:buncher-ycf-identified}.

    \item  $ \widehat\tau_S$  compares the average treated outcome for shifters in the window $X^{kp}\in (x^*,x^{kp,out}]$ with their average counterfactual outcome in the corresponding interval $X^{cf} \in (x^*+\widehat{\Delta x}^{\,*},\widehat x^{cf,out}]$, based on equation \eqref{eq:tau-s-diffuse}.
    \end{itemize} 

\end{enumerate} 

\subsection{Inference, Reporting and Sensitivity}
\label{sec:inference} \label{sec:reporting-sensitivity}

Inference is based on bootstrap that re-estimates the entire procedure. Each bootstrap draw resamples observations from the focal and placebo groups, recomputes the observed densities and outcome functions, refits \(\widehat\delta_h\) and \(\widehat\delta_y\), re-estimates \(\widehat h_F^{cf}(x)\), \(\widehat B_F\), \(\widehat{\Delta x}^{\,*}\), \(\widehat x^{cf,out}\), \(\widehat m_F^{cf}(x)\), and recalculates the aggregate effects  \(\widehat\tau_B\), \(\widehat\tau_S\).\footnote{When observations are clustered, the bootstrap resamples at the level of the independent sampling unit. When the placebo group is drawn from a different population, time period, or policy regime, the bootstrap resamples the focal and placebo samples separately. The local window, bin width or bandwidth, basis functions, and comparison group are held fixed in the baseline bootstrap. } Standard errors and confidence intervals are computed from the bootstrap distribution of the final estimators.  

The empirical section should report three sets of objects. First, it should report the assignment-side objects: the observed densities, the estimated counterfactual density, the excess bunching mass, the estimated marginal response \(\widehat{\Delta x}^{*}\),  and the equal-mass endpoint for shifters. Second, it should report the outcome-side objects: the observed outcome functions, the fitted counterfactual outcome function, the aggregate treatment effects for shifters and bunchers. Third, it should report sensitivity to the local window, bin width, placebo group, density and outcome specifications, and estimates at pseudo thresholds. It is the roadmap for the application below.


\section{Application: Kinked Coinsurance Policy in China}
\label{sec:application}

%
%

We apply the framework to China's kinked outpatient coinsurance schedule. The policy reimburses eligible annual expenditure at a low copayment rate up to a cap $x^*$ and at the full marginal cost above it---a convex kink that induces bunching at $x^*$. URRBMI enrollees are the focal group $F$; contemporaneous UEBMI enrollees, whose caps are far higher so that $x^*$ is not a kink for them, are the placebo group $P$. Using the UEBMI shape together with URRBMI's own unaffected region below $x^*$ to discipline the level-and-slope adjustments, we recover the counterfactual assignment density $h_F^{cf}$ and marginal response $\Delta x^*$, the counterfactual outcome function $m_F^{cf}$, and the average effects on bunchers and shifters, $\tau_B$ and $\tau_S$.

\subsection{Institutional Setting}
\label{sec:app-institutions}

China's basic medical insurance system is organized around two programs that together cover the near-universe of the population. The Urban Employee Basic Medical Insurance (UEBMI) covers formal-sector employees and is financed by payroll contributions shared between employers and employees. The Urban and Rural Residents Basic Medical Insurance (URRBMI) covers the remaining population---rural residents and non-employed urban residents such as children, students, and the unemployed---and is financed by enrollee premiums together with substantial government subsidies.\footnote{UEBMI premiums are set as a share of the previous year's wage, with employers contributing roughly $6\%$ and employees roughly $2\%$. URRBMI premiums are largely subsidized by the government, with enrollees paying a small fixed amount.} URRBMI was created by merging the former rural scheme (NRCMS) and the urban-resident scheme (URBMI) into a single program. By 2011 the two programs jointly covered more than $92\%$ of the urban and $97\%$ of the rural population \citep{yu2015universal}.

Both programs reimburse outpatient care through a piecewise-linear cost-sharing schedule defined over annual \emph{eligible} medical expenditure, that is, spending on services and drugs within the insurance catalog. This eligible expenditure is the assignment variable \(X\) of our framework. Up to an annual reimbursement threshold \(x^*\), the enrollee pays a coinsurance share \(\tau\) of each additional yuan of eligible spending; once cumulative eligible spending exceeds \(x^*\), reimbursement stops and the enrollee bears the full marginal cost. Annual out-of-pocket payments are therefore
\[
\mathrm{OOP}(X)=
\begin{cases}
\tau\,X, & X\le x^*,\\[0.3em]
\tau\,x^* + (X-x^*), & X> x^*.
\end{cases}
\]
This is precisely the kinked schedule \(T^{kp}\) of Section~\ref{sec:policy-environment}, with below-threshold marginal rate \(t=\tau\) and above-threshold marginal rate \(t+\Delta t=1\), so that the kink has size \(\Delta t=1-\tau\) at \(x^*\). Because the schedule is convex at \(x^*\), enrollees whose latent eligible expenditure lies just above the threshold have an incentive to reduce spending toward \(x^*\); this generates bunching at the threshold and induces shifting among enrollees who remain above it.

For URRBMI, the annual ceiling was \(x^*=600\) RMB in 2011 and \(x^*=800\) RMB in 2012, with coinsurance rates of $50\%$ at community health centers and $60\%$ at hospitals.\footnote{Under UEBMI, ceilings are far higher and additionally depend on employment status (active versus retired) and on whether the condition is chronic. We summarize the full plan parameters of both programs in the data appendix.} UEBMI applies much higher ceilings and lower coinsurance rates, so neither $600$ nor $800$ RMB constitutes a kink for UEBMI enrollees. This institutional contrast supplies the two sources of variation that our design exploits. First, contemporaneous UEBMI enrollees serve as a placebo group \(P\) whose density and outcome profiles are smooth through the URRBMI ceilings; they discipline the shape of the URRBMI counterfactual through the design-assisted restrictions of Sections~\ref{sec:density-shape-restriction} and~\ref{sec:outcome-shape-restriction}. Second, because the URRBMI ceiling moved from $600$ to $800$ RMB between the two years, each year's non-binding threshold---$800$ RMB in 2011 and $600$ RMB in 2012---provides a within-group placebo that we use in the validation tests of Section~\ref{sec:app-robustness}. Throughout the application we take URRBMI as the focal group \(F\), using 2012 (ceiling $800$ RMB) as the baseline case and 2011 (ceiling $600$ RMB) as a complementary case.

\subsection{Data and Analysis Sample}
\label{sec:app-data}

Our analysis draws on the universe of outpatient insurance claims from a prefecture in eastern China for 2011 and 2012. The URRBMI data covers approximately 19 million visits in 2011 and 21 million visits in 2012, including details in medical expenses.  Claims are filed directly by hospitals and pharmacies to the insurer, so each visit generates a record automatically and dispensed drugs are billed to the same account. We aggregate these visit-level records to the enrollee--year, which yields each enrollee's annual eligible expenditure---the assignment variable \(X\)---together with annual measures of utilization. Because \(X\) is built from administrative claims rather than self-reports, the bunching documented below reflects a real adjustment of eligible spending rather than relabeling or misreporting, which rules out the reporting-response caveat of Section~\ref{sec:relabeling-reference-points}.\footnote{Over-the-counter purchases made entirely outside the insurance system are not recorded and fall outside \(X\); these are typically small expenditures for minor conditions that do not require a prescription.}

To trace the margins through which the kink operates, we examine a set of ouctomes measured at the enrollee--year level: (1) the total number of annual outpatient visits (\texttt{novisit}), (2) average cost per visit (\texttt{zfy\_avg}),  and 
(3) the composition of care: whether the visit is at a hospital or community health center, measured by hospital-visit share (\texttt{jglb\_hospital}), and the number of hospital visits (\texttt{no\_hospital}); 
(c)  the last visit time, measured by the month and quarter of  last visit (\texttt{last\_m}, \texttt{last\_q}). 
Together these capture not only how much care enrollees use, but also its intensity and  composition. 

\subsection{Bunching and Stylized Facts}
\label{sec:app-stylized}

Before estimating any counterfactual, we present two pieces of model-free evidence that motivate the design. Both track the year-specific URRBMI threshold and are absent for UEBMI: a spike in the assignment density at the kink, and a discontinuity in outpatient utilization at the same point.

\subsubsection{Bunching in the assignment density}
\label{sec:app-bunching}

\begin{figure}[htbp]
\centering
\caption{Density of eligible annual medical expenses, URRBMI and UEBMI, 2011 and 2012.}
\includegraphics[width=0.7\linewidth]{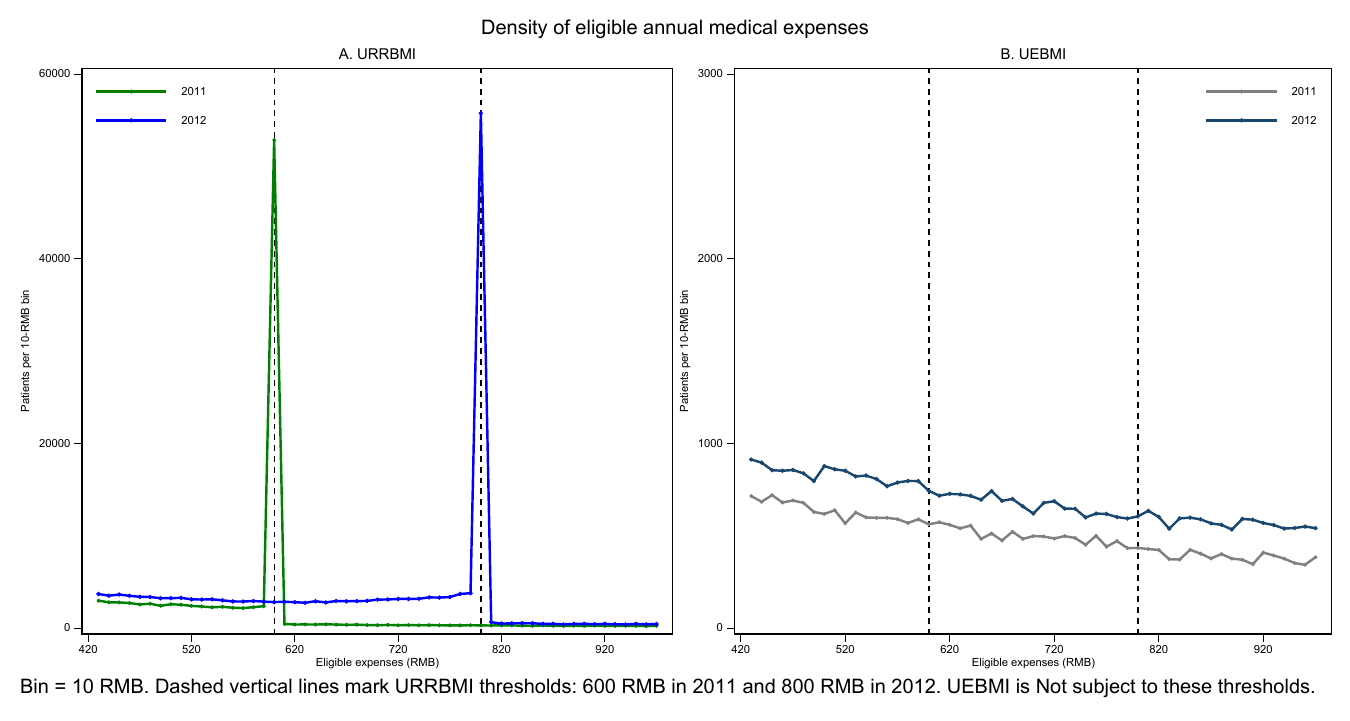}
\begin{minipage}{16cm} 
\footnotesize{ Panel~A: URRBMI bunches sharply at its year-specific reimbursement thresholds ($600$ RMB in 2011, $800$ RMB in 2012), with a deficit just above. Panel~B: UEBMI, with much higher thresholds, is smooth through both thresholds. Bin width $10$ RMB; the vertical axis counts enrollees per bin; dashed vertical lines mark $600$ and $800$ RMB.}
\end{minipage}
\label{fig:app-raw-density}
\end{figure}

Figure~\ref{fig:app-raw-density} plots the number of enrollees per $10$-RMB bin of eligible expenditure \(X\), separately for 2011 and 2012. Panel A (URRBMI) shows pronounced excess mass exactly at the year's reimbursement ceiling: a spike at $600$ RMB in 2011 and at $800$ RMB in 2012, each accompanied by a visible deficit of mass just above the threshold. The location of the bunching tracks the policy--there is no spike at $800$ RMB in 2011 or at $600$ RMB in 2012, when those values are not kinks.   
Panel B (UEBMI), whose ceilings are far higher, is smooth through both $600$ and $800$ RMB, confirming that the excess mass is a response to the URRBMI kink rather than an artifact of how expenditure accumulates. This excess mass is the bunching \(B_F\) that the mass-balance condition~\eqref{eq:B-M} links to the counterfactual density and the marginal response \(\Delta x^*\).  

\subsubsection{Stylized facts on outpatient utilization}
\label{sec:app-stylized-outcome}

\begin{figure}[htbp]
\centering
\caption{Annual outpatient visits.} 
\includegraphics[width=0.7\linewidth]{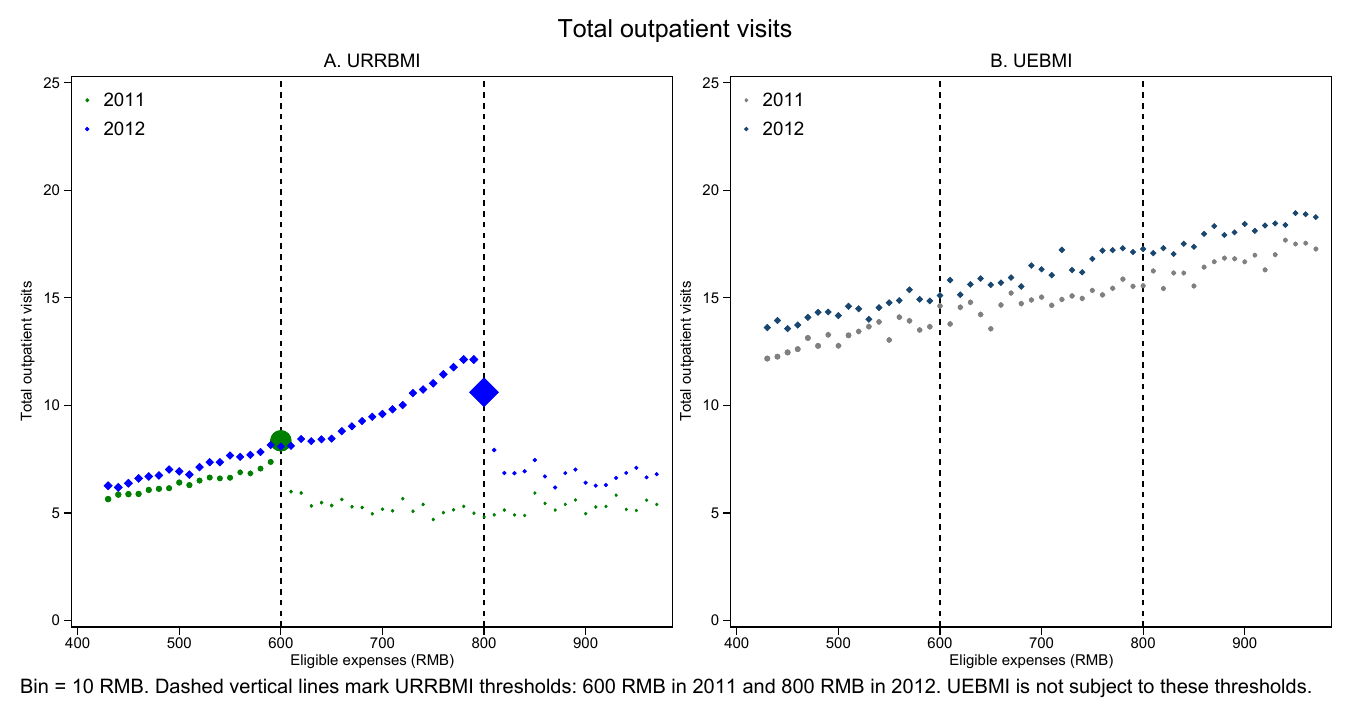}
\begin{minipage}{16cm}
\footnotesize{Note: Panel~A: URRBMI visits rise below the threshold, drop sharply at the year-specific threshold, and flatten above it. Panel~B: UEBMI is smooth through $600$ and $800$ RMB, as they are not subject to these thresholds. Markers are bin-level means weighted by bin counts; bin width $10$ RMB.} 
\end{minipage}
\label{fig:app-raw-outcome}
\end{figure}

Figure~\ref{fig:app-raw-outcome} plots the bin-level mean of annual outpatient visits against eligible expenditure, weighted by bin size. For URRBMI (Panel A), number of visits rise with expenditure below the threshold, drop sharply at the year-specific threshold, and flatten above it---the pattern expected if the jump to full marginal cost above \(x^*\) discourages further utilization. As with the density, the discontinuity appears only at the binding threshold ($600$ RMB in 2011, $800$ RMB in 2012) and not at the non-binding threshold; for UEBMI (Panel B) the visit profile is smooth through both values.

\begin{figure}[htbp]
\centering
\caption{Average cost per outpatient visit.} 
\includegraphics[width=0.7\linewidth]{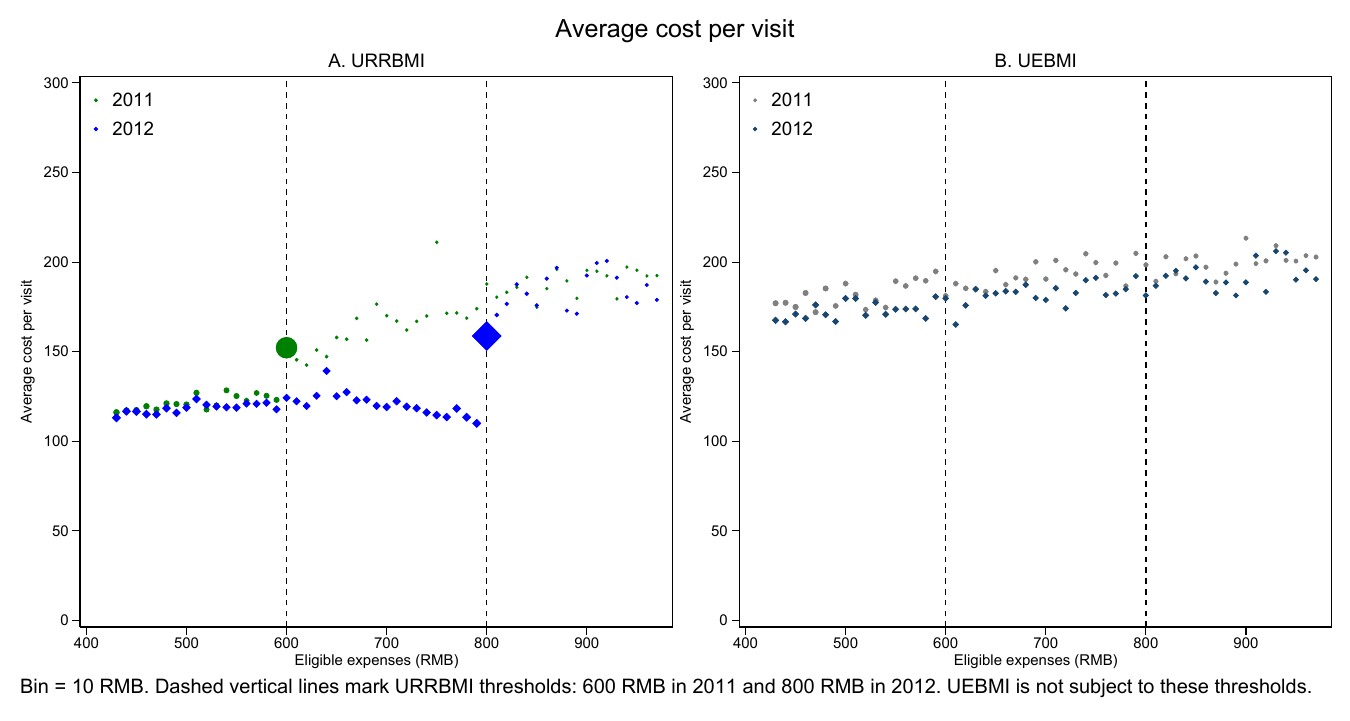}
\begin{minipage}{16cm}
\footnotesize{Note: Panel~A: URRBMI cost per visit is flat below the year-specific threshold and jumps upward above it. Panel~B: UEBMI rises smoothly through $600$ and $800$ RMB, as it is not subject to these thresholds. Markers are bin-level means weighted by bin counts; bin width $10$ RMB.} 
\end{minipage}
\label{fig:app-raw-cost}
\end{figure}

The threshold reshapes the \emph{intensity} of care as well. Figure~\ref{fig:app-raw-cost} plots the bin-level mean cost per outpatient visit. For URRBMI (Panel~A), cost per visit is flat at around $120$~RMB below the year-specific threshold and jumps upward just above it, to roughly $180$--$200$~RMB; for UEBMI (Panel~B) it rises smoothly with expenditure and shows no break at $600$ or $800$~RMB. Read alongside the visit profile, the two facts describe enrollees above the cap make fewer visits but spend more on each visit, indicating a shift in the composition of care toward more serious, higher-cost conditions once the reimbursement ceiling binds.

These are descriptive facts, not treatment effects. The observed profile compares \emph{different} enrollees at each expenditure level, and around the kink the observed population is reshaped by bunching and shifting: enrollees seen just above \(x^*\) are not the ones who would have chosen that level absent the kink.  Recovering causal effects therefore requires constructing the counterfactual density and outcome function, and then comparing treated outcomes with counterfactual outcomes for the affected populations: bunchers over their missing counterfactual interval and shifters over the equal-mass interval corresponding to the chosen observed shifter window.


\subsection{Design-Assisted Counterfactual Assignment Density}
\label{sec:app-density}
We now implement the design-assisted density restriction of Section~\ref{sec:density-shape-restriction} for the 2012 URRBMI cohort, taking UEBMI as the placebo group $P$ and the $800$~RMB ceiling as the kink $x^*$. The object of interest is the focal counterfactual density $h_F^{cf}$, which fixes the marginal bunching response $\Delta x^*$ and, through it, the partition of enrollees into the unaffected, buncher, and shifter groups carried into the outcome analysis in Section~\ref{sec:app-outcomes}.

Following Assumption~\ref{assume:density-shape} we take $q_h(x)=(1,x)'$, so URRBMI and UEBMI may differ in both the level and the local slope of their densities while any higher-order curvature is treated as common to the two groups. The adjustment $\delta_h$ is identified on URRBMI's unaffected region below the cutoff---bins in $[420,760)$, excluding the diffuse bunching window---via Equation \eqref{eq:delta-h-identification}; 
combined with the UEBMI density it delivers $h_F^{cf}$ on the missing region above $x^*$ through Equation~\eqref{eq:hcf-identified}. The marginal response then solves the buncher mass-balance condition $B_F=M(\Delta x^*)$ in Equation~\eqref{eq:delta-mass-balance}: $\Delta x^*$ is the width of the interval above $x^*$ whose counterfactual mass equals the observed excess bunching. Given the observed evaluation endpoint $x^{kp,out}=980$~RMB, the equal-mass condition~\eqref{eq:xcf-shift-end}--\eqref{eq:xcf-all-end} determines the corresponding counterfactual endpoint $x^{cf,out}$.

\begin{figure}[htbp]
\centering
\caption{Observed and Counterfactual Density, URRBMI 2012 (baseline Poisson/log-link cubic).}
\includegraphics[width=0.6\linewidth]{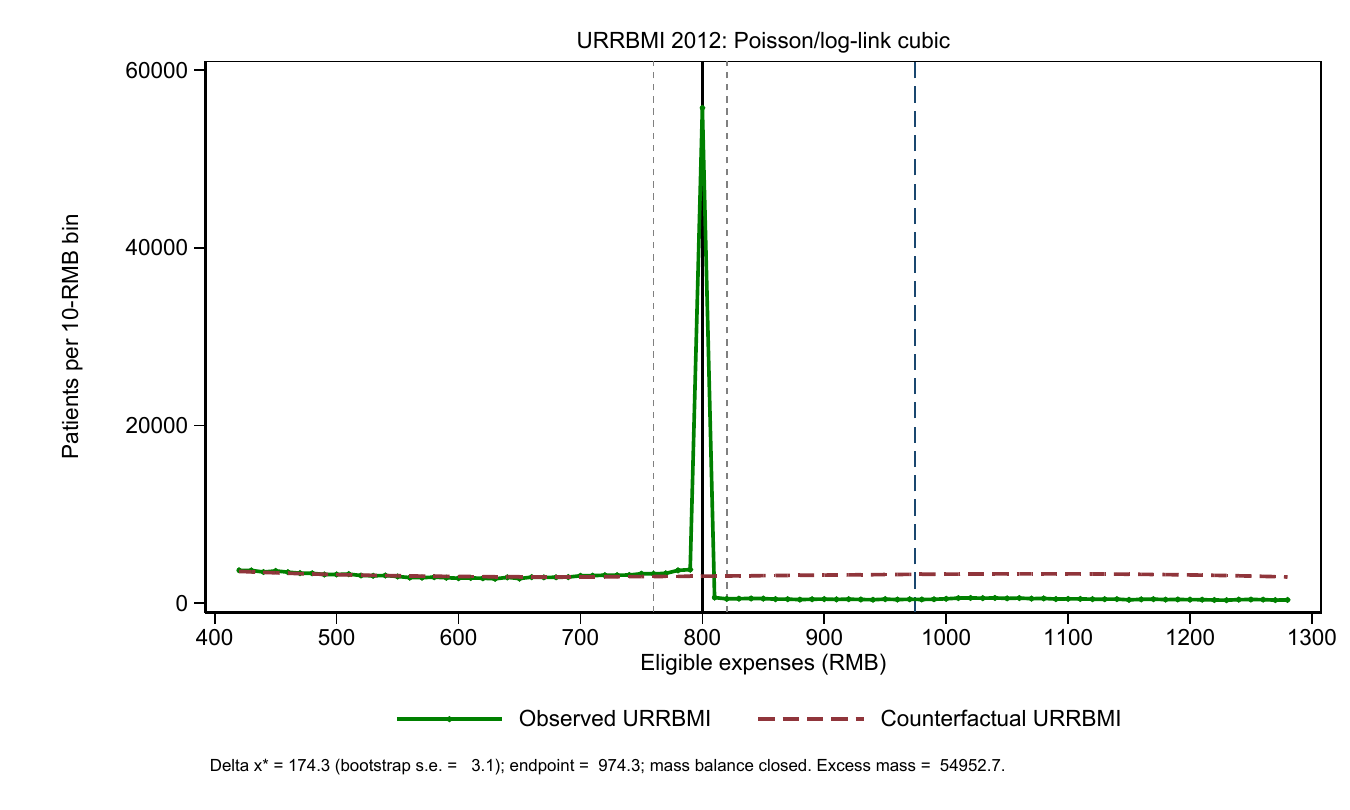} 
\begin{minipage}{16cm}
\footnotesize{Note: The solid line is the observed URRBMI density (enrollees per $10$-RMB bin); the dashed line is the design-based counterfactual $h_F^{cf}$, identified from the UEBMI placebo density and the level-and-slope adjustment $\delta_h$ fit on URRBMI's unaffected region below the cutoff. The solid vertical line marks the kink $x^*=800$~RMB; the inner dashed verticals mark the diffuse window $[760,820]$ excluded from the unaffected fit. Baseline $\widehat{\Delta x}^{*}=174.3$~RMB (bootstrap s.e.\ $3.1$).}  
\end{minipage}
\label{fig:app-cf-density}
\end{figure}

Figure~\ref{fig:app-cf-density} reports the baseline estimate (Poisson/log-link cubic, estimation window $[420,1280]$, diffuse window $[760,820]$, $10$-RMB bins). The dashed curve is the design-based counterfactual density. 
The marginal bunching response is $\widehat{\Delta x}^{*}=174.3$~RMB (bootstrap s.e.\ $3.1$, $95\%$ percentile CI $[168.2,180.2]$), so the counterfactual location of the marginal buncher is $x^*+\widehat{\Delta x}^{*}=974.3$~RMB, indicating the marginal buncher's eligible expenditure would have been about $1.2$ times the kink value.
The estimate partitions enrollees into the three behavioral groups used in Section~\ref{sec:app-outcomes}: unaffected ($x^{cf}\le 800$), bunchers ($x^{cf}\in(800,974.3]$), and shifters ($x^{cf}>974.3$).  
 The reported shifter effect below is evaluated on the observed window ending at \(x^{kp,out}=980\)~RMB and its equal-mass counterfactual interval. 

The estimate is stable across the density specification. Table~\ref{tab:app-deltax} reports $\widehat{\Delta x}^{*}$ for six choices of the basis $q_h$ and link function. The positively-constrained count specifications cluster between $174$ and $189$~RMB, with the level (OLS) cubic highest at $198$~RMB; every bootstrap percentile interval is tight and excludes zero, and all specifications close mass balance.
 The negative-binomial cubic gives the best fit to the UEBMI density (RMSE $26.9$ against $35.8$ for the baseline) while returning a somewhat larger $\widehat{\Delta x}^{*}=188.7$~RMB. We retain the Poisson/log-link cubic as the baseline because it is the most conservative of the count specifications and the simplest, and carry the negative binomial as a validation-oriented robustness check in Section~\ref{sec:app-robustness}.

\begin{table}[htbp]
\centering
\caption{Marginal bunching response $\widehat{\Delta x}^{*}$ by counterfactual-density specification, URRBMI 2012.}
\label{tab:app-deltax}
\begin{tabular}{lccc}
\toprule
Specification & $\widehat{\Delta x}^{*}$ (RMB) & Bootstrap s.e. & $95\%$ CI  \\  
\midrule
Poisson/log-link cubic (baseline) & $174.3$ & $3.1$ & $[168.2,\,180.2]$ \\  
Poisson $\log(x)$ cubic           & $177.3$ & $3.0$ & $[171.6,\,182.9]$  \\ 
Poisson spline                    & $174.7$ & $3.4$ & $[167.9,\,181.0]$  \\ 
Poisson reciprocal cubic          & $185.7$ & $3.1$ & $[179.5,\,192.2]$  \\ 
Negative binomial cubic           & $188.7$ & $3.7$ & $[182.6,\,197.5]$  \\ 
Level polynomial cubic (OLS)      & $198.3$ & $2.9$ & $[192.6,\,203.4]$  \\ 
\bottomrule
\end{tabular}
\begin{minipage}{16cm}
\vskip 4pt
\footnotesize{Note: All specifications use estimation window $[420,1280]$, diffuse window $[760,820]$, $10$-RMB bins, and $300$ bootstrap draws resampling URRBMI and UEBMI individuals separately within the window. $\widehat{\Delta x}^{*}$ solves the buncher mass-balance condition~\eqref{eq:delta-mass-balance}; the $95\%$ CI is the bootstrap percentile interval. UEBMI RMSE is the in-window root-mean-square fit error of the placebo density and indexes specification quality (lower is better). All specifications close mass balance with every bootstrap draw valid.}
\end{minipage}
\end{table}

\subsection{Counterfactual Outcomes and Treatment Effects}
\label{sec:app-outcomes}

We now turn from the assignment density to outcomes. 
The counterfactual outcome function is disciplined by the same placebo logic used for the density: $m_F^{cf}(x)=m_P(x)+q_y(x)'\delta_y$ (Equation~\eqref{eq:outcome-shape}), with $\delta_y$ identified on URRBMI's unaffected region below the cutoff (Equations~\eqref{eq:unaffected-outcome-equality}--\eqref{eq:delta-y-identification}), giving $m_F^{cf}$ on the missing region (Equation~\eqref{eq:mcf-identified}). Based on the estimated counterfactual density $h_F^{cf}$ and counterfactual outcome function $m_F^{cf}$, we can compute the   average effect on bunchers $\tau_B$ (Equation~\eqref{eq:tau-b-identified}) 
 and on shifters $\tau_S$ (Equation~\eqref{eq:tau-s-identified}), 
with inference from $300$ bootstrap estimates. 

\begin{figure}[htbp]
\centering
\caption{Observed and counterfactual outpatient visits, URRBMI 2012 (baseline specification).} 
\includegraphics[width=0.6\linewidth]{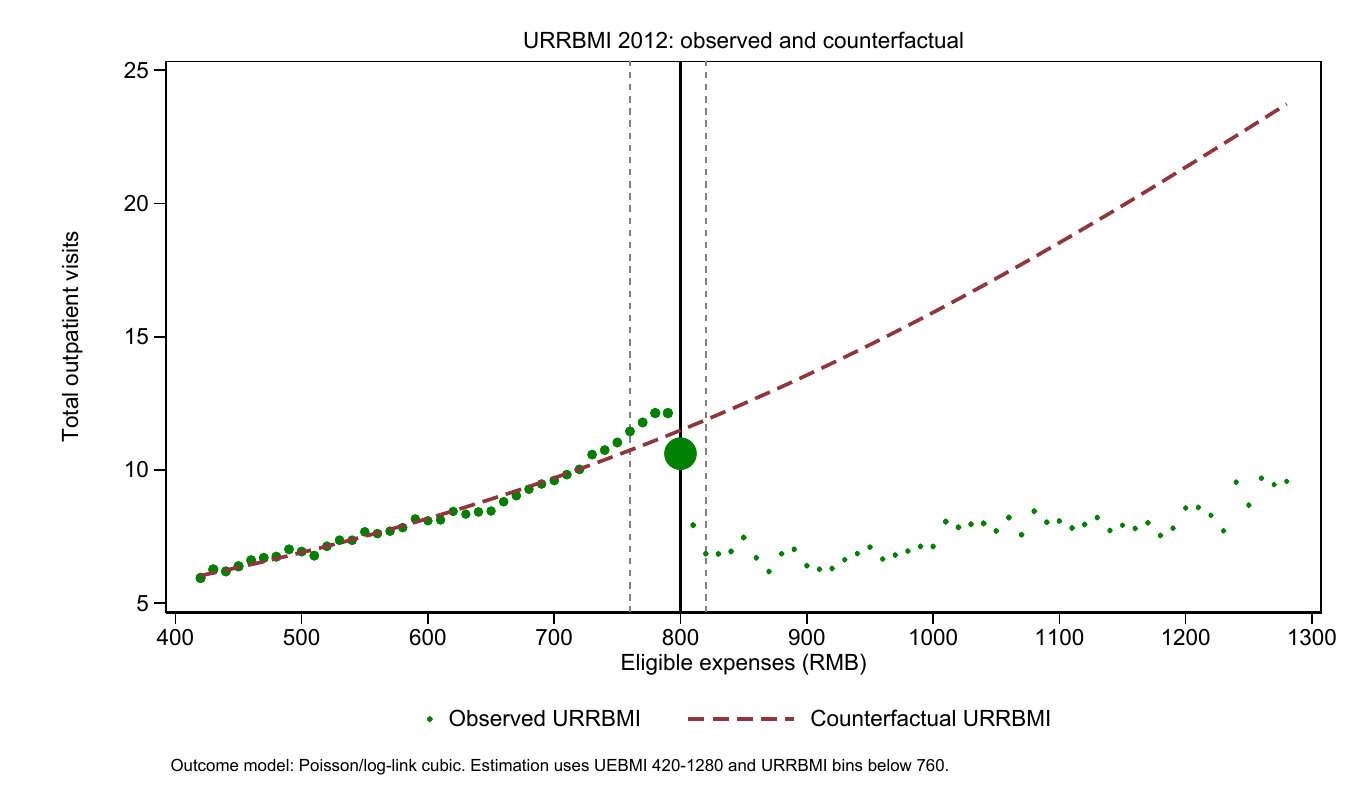}
\begin{minipage}{16cm}
\footnotesize{Note: Green markers are bin-level mean outpatient visits for observed URRBMI; the dashed line is the design-based counterfactual outcome function $m_F^{cf}$, identified from the UEBMI visit profile and the adjustment $\delta_y$ fit on URRBMI bins below the cutoff. The solid vertical marks the kink $x^*=800$~RMB; inner dashed verticals mark the diffuse window $[760,820]$. 
Outcome model Poisson/log-link cubic; estimation uses UEBMI $[420,1280]$ and URRBMI bins below $760$.}
\end{minipage}
\label{fig:app-cf-outcome}
\end{figure}

 Figure~\ref{fig:app-cf-outcome} shows the construction for the annual number of outpatient visits (\texttt{novisit}). Observed visits rise with eligible expenditure up to the threshold; the dashed counterfactual curve continues to rise steeply above it---this is the utilization that capped enrollees forgo. The treatment effects are interval averages for affected populations:
 bunchers collapse from $X^{cf}\in(800,974]$ onto $X^{kp}=800$, while the observed shifters window $X^{kp}\in(800,980]$ originated from the counterfactual interval $X^{cf}\in(974,999.3]$ via equal-mass condition.\footnote{The observed shifter endpoint $x^{kp,out}=980$ selects the range of shifters under study; the counterfactual endpoint $x^{cf,out}=999.3$ is then pinned down by the equal-mass condition for shifters. The choice of $980$ is not essential and can be widened or narrowed, depending on the shifter population being studied.}

At the baseline specification, \textit{bunchers} reduce their annual number of outpatient visits by $\widehat\tau_B=-2.58$ ($95\%$ CI $[-2.86,-2.28]$), compared to their counterfactual; i.e., observed $10.8$ vs. counterfactual $13.4$ number of visits. \textit{Shifters} reduce their annual outpatient visits by $\widehat\tau_S=-8.92$   ($95\%$ CI $[-9.35,-8.48]$), , compared to their counterfactual; i.e.,  observed $6.7$ vs.\ counterfactual $15.7$. Both effects are large and precisely estimated. 
The estimates are stable across the six density specifications, with $\widehat\tau_B\in[-2.84,-1.93]$ and $\widehat\tau_S\in[-9.48,-7.73]$.

\begin{figure}[htbp]
\centering
\caption{\small{Observed and counterfactual average cost per visit, URRBMI 2012 (baseline specification).}}
\includegraphics[width=0.6\linewidth]{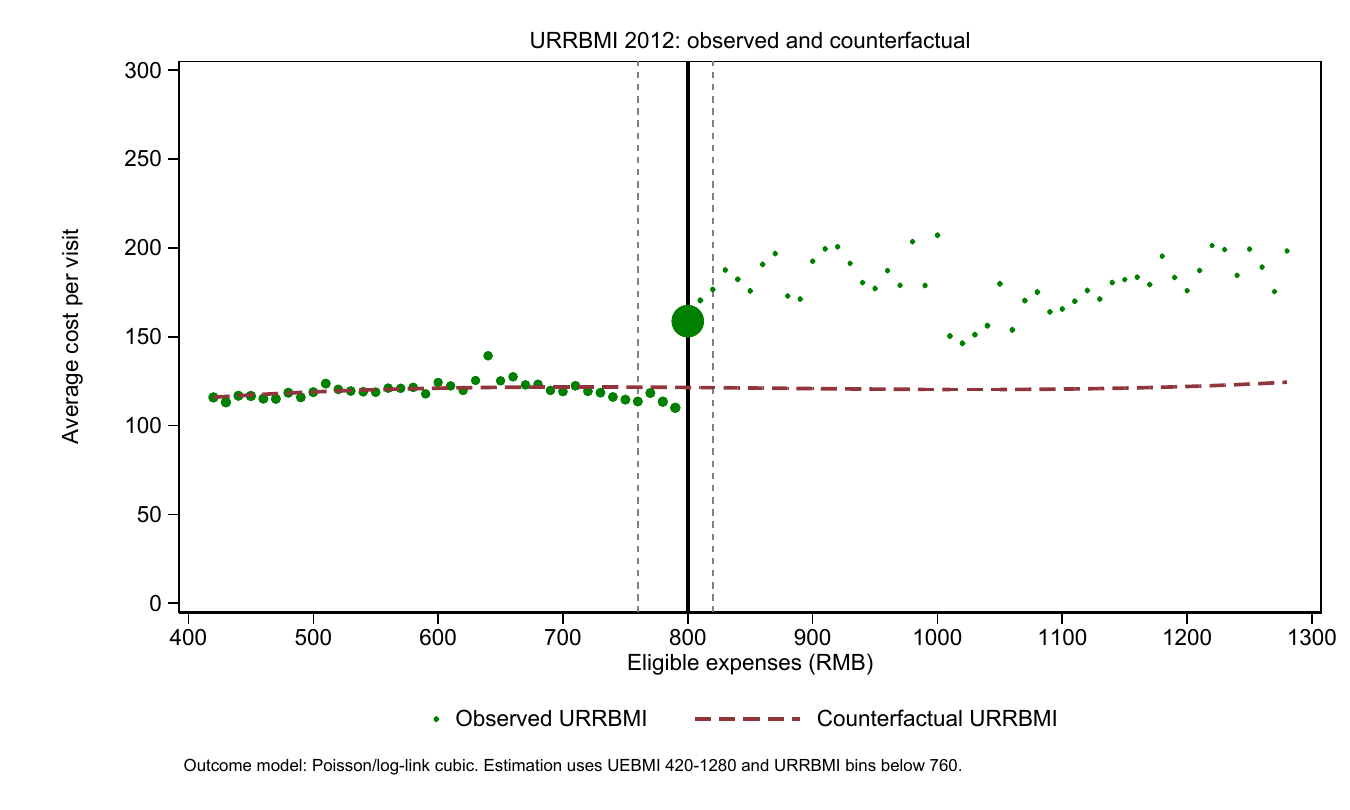}
\begin{minipage}{16cm}
\footnotesize{Note: Green markers are bin-level mean cost per outpatient visit (RMB) for observed URRBMI; the dashed line is the design-based counterfactual outcome function $m_F^{cf}$, identified from the UEBMI cost profile and the adjustment $\delta_y$ fit on URRBMI bins below the cutoff. The solid vertical marks the kink $x^*=800$~RMB; inner dashed verticals mark the diffuse window $[760,820]$. Above the cap the observed cost per visit rises well above the counterfactual, consistent with enrollees concentrating their remaining visits on more serious, higher-cost conditions. Outcome model Poisson/log-link cubic; estimation uses UEBMI $[420,1280]$ and URRBMI bins below $760$.}
\end{minipage}
\label{fig:app-cf-cost}
\end{figure}

The visit reduction coincides with a sharp rise in the \emph{average cost per visit} (\texttt{zfy\_avg}),  shown in Figure~\ref{fig:app-cf-cost}. Below the cap, observed and counterfactual cost per visit coincide at around $120$~RMB; above it, the observed cost per visit jumps well above the smooth counterfactual. Bunchers pay $\widehat\tau_B=+36.3$~RMB more per visit than their counterfactual ($95\%$ CI $[32.4,40.1]$) and shifters pay  $\widehat\tau_S=+66.8$~RMB more ($[60.3,73.2]$), an increase of roughly $55\%$.

Fewer visits, each more expensive, reflects a shift in the \emph{composition} of care once the subsidy is exhausted. The reduction is not uniform across providers: enrollees cut back on low-cost community health-center visits while their use of hospital care, if anything, rises. The hospital share of visits increases by $\widehat\tau_S=+0.38$ and the number of hospital visits by $\widehat\tau_S=+1.40$, so the entire decline in total visits is borne by community health-center care.\footnote{The observed and design-based counterfactual profiles for the hospital visit share and the number of hospital visits are reported in Figures~\ref{fig:app-cf-hosp-share} in Appendix~\ref{app:additional}.} 
Once facing the full marginal cost of care, the enrollees forgo minor diseases and reserve out-of-pocket spending for more serious conditions, which mechanically raises the average cost of the visits that remain. 

Care also stops earlier in the year. The calendar month of the last outpatient visit falls by about $0.9$ of a month for shifters and $0.6$ for bunchers (Figure~\ref{fig:app-cf-last-m} in Appendix~\ref{app:additional}). Absent the policy, the timing of an enrollee's final illness should vary smoothly with expenditure. 
So, a discrete downward shift at and above the threshold is hard to attribute to  \emph{when} people fall ill;
 it instead reveals capped enrollees who stop seeking care for illnesses that arise after their reimbursement is exhausted. The same pattern holds when timing is measured by the calendar quarter of the last visit.

 Table~\ref{tab:app-effects} collects the baseline effects across the utilization, cost, composition, and timing margins. The effects are concentrated among shifters; buncher effects are smaller throughout, as expected since bunchers move only a short distance, and all are stable across the six density specifications.

\begin{table}[htbp]
\centering
\caption{Treatment effects on bunchers and shifters, URRBMI 2012 (baseline specification).}
\label{tab:app-effects}
\small
\begin{tabular}{l ccc ccc}
\toprule
 & \multicolumn{3}{c}{Bunchers} & \multicolumn{3}{c}{Shifters} \\
\cmidrule(lr){2-4}\cmidrule(lr){5-7}
Outcome &$\bar{Y}_B^{cf}$ & $\widehat\tau_B$  & $95\%$ CI of $\widehat\tau_B$ &$\bar{Y}_S^{cf}$  &$\widehat\tau_S$ & $95\%$ CI of $\widehat\tau_S$\\
\midrule
Total outpatient visits& $13.39$ & $-2.58$   &$[-2.86,-2.28]$ & $15.66$  & $-8.92$  &$[-9.35,-8.48]$ \\
Average cost per visit & $120.9$ & $36.3$  &$[32.4,\ 40.1]$   & $120.4$ &$66.8$  &$[60.3,\ 73.2]$  \\
\addlinespace
\multicolumn{7}{l}{\textit{Composition}} \\
\quad Hospital visit share & $0.137$ &$-0.012$   &$[-0.021,-0.005]$ & $0.134$ &$0.377$   &$[\ 0.364,\ 0.389]$  \\
\quad Hospital visits & $0.87$ & $0.03$   &$[-0.03,\ 0.09]$ & $0.97$  & $1.40$   &$[\ 1.29,\ 1.48]$  \\
\addlinespace
\multicolumn{7}{l}{\textit{Timing}} \\
\quad Last visit (month)& $10.03$ & $-0.58$   &$[-0.67,-0.49]$ & $10.10$ & $-0.91$   &$[-1.04,-0.76]$  \\
\quad Last visit (quarter) & $3.56$ & $-0.16$   &$[-0.19,-0.14]$ & $3.59$ & $-0.28$   &$[-0.32,-0.24]$  \\
\bottomrule
\end{tabular}
\begin{minipage}{16cm}
\vskip 4pt
\footnotesize{Note: Baseline specification (Poisson/log-link cubic density, estimation window $[420,1280]$, diffuse window $[760,820]$, $10$-RMB bins). $\widehat\tau_B$ is the average effect on bunchers, whose counterfactual interval is $X^{cf}\in(x^*,x^*+\Delta x^*]$, based on equations~\eqref{eq:tau-b-identified}. 
 $\widehat\tau_S$ is the average effects on shifters in the evaluation window $X^{kp}\in (x^*, x^{kp,out}]$, compared with its equal-mass counterfactual interval $X^{cf}\in (x^*+\Delta x^*, x^{cf,out}]$ , based on equation ~\eqref{eq:tau-s-identified}. $\bar{Y}_B^{cf}$ and $\bar{Y}_S^{cf}$ are the corresponding average counterfactual outcome means, so that $\widehat\tau=\bar Y^{kp}-\bar Y^{cf}$.
$95\%$ CI is the bootstrap confidence interval across $300$ bootstrap draws that resample URRBMI and UEBMI individuals separately.
``Average cost per visit'' is in RMB; ``hospital visits'' is a count; ``hospital visit share'' is a probability; ``last visit (month/quarter)'' is the calendar month or quarter of the enrollee's final outpatient visit.}
\end{minipage}
\end{table}

\subsection{Robustness and Design Validation}
\label{sec:app-robustness}
A central advantage of the design-based approach over treated-only extrapolation is that its identifying restrictions have testable implications: the counterfactual is disciplined by an observable placebo group rather than by a functional form imposed on the treated data alone. We report two kinds of evidence---the stability of the estimated effects across specifications, and a placebo test at a non-binding threshold, and .

\textbf{Specification robustness of the treatment effects.} Section~\ref{sec:app-density} showed that the marginal bunching response $\widehat{\Delta x}^{*}$ is stable across density specifications; the estimated treatment effects inherit that stability. Table~\ref{tab:app-robustness} reports average treatment effect on \textit{bunchers} $\widehat\tau_B$ and average effect on \textit{shifters} $\widehat\tau_S$ for the two primary outcomes across all six specifications. For total number of outpatient visits, the shifter effect ranges over $[-9.48,-7.73]$ and the buncher effect over $[-2.84,-1.93]$. For average cost per visit, $\widehat\tau_S\in[62.3,69.0]$ and $\widehat\tau_B\in[34.5,37.6]$. The level (OLS) specification, which returns the largest $\widehat{\Delta x}^{*}$, is the mild outlier with smaller visit effects, while the five count specifications cluster tightly. Every effect retains its sign and statistical significance throughout.

\begin{table}[htbp]
\centering
\caption{Treatment effects by specification, URRBMI 2012 (primary outcomes).}
\label{tab:app-robustness}
\small
\begin{tabular}{l c cc cc}
\toprule
 & & \multicolumn{2}{c}{Total Visits } & \multicolumn{2}{c}{Cost per Visit } \\
\cmidrule(lr){3-4}\cmidrule(lr){5-6}
Specification & $\widehat{\Delta x}^{*}$ & $\widehat\tau_B$ & $\widehat\tau_S$ & $\widehat\tau_B$ & $\widehat\tau_S$ \\
\midrule
Poisson/log-link cubic (baseline) & $174.3$ & $-2.58$ & $-8.92$ & $36.3$ & $66.8$ \\
                                  & $(3.1)$ & $(0.14)$ & $(0.21)$ & $(2.0)$ & $(3.1)$ \\
\addlinespace
Poisson $\log(x)$ cubic           & $177.3$ & $-2.66$ & $-9.01$ & $36.2$ & $65.3$ \\
                                  & $(3.0)$ & $(0.14)$ & $(0.20)$ & $(1.8)$ & $(2.9)$ \\
\addlinespace
Poisson spline                    & $174.7$ & $-2.57$ & $-8.67$ & $36.4$ & $63.4$ \\
                                  & $(3.4)$ & $(0.14)$ & $(0.21)$ & $(2.0)$ & $(3.4)$ \\
\addlinespace
Poisson reciprocal cubic          & $185.7$ & $-2.71$ & $-9.18$ & $34.5$ & $62.3$ \\
                                  & $(3.1)$ & $(0.14)$ & $(0.21)$ & $(1.9)$ & $(3.0)$ \\
\addlinespace
Negative binomial cubic           & $188.7$ & $-2.84$ & $-9.48$ & $37.6$ & $69.0$ \\
                                  & $(3.7)$ & $(0.15)$ & $(0.23)$ & $(2.0)$ & $(3.1)$ \\
\addlinespace
Level polynomial cubic (OLS)      & $198.3$ & $-1.93$ & $-7.73$ & $36.9$ & $68.0$ \\
                                  & $(2.9)$ & $(0.13)$ & $(0.18)$ & $(2.4)$ & $(3.7)$ \\
\bottomrule
\end{tabular}
\begin{minipage}{15cm}
\vskip 4pt
\footnotesize{Note: Each row is one density/outcome specification (estimation window $[420,1280]$, diffuse window $[760,820]$, $10$-RMB bin); bootstrap standard errors in parentheses. $\widehat{\Delta x}^{*}$ is the marginal bunching response in RMB (cf.\ Table~\ref{tab:app-deltax}). $\widehat\tau_B$ and $\widehat\tau_S$ are the average effects on bunchers and shifters; total visits are counts and cost per visit is in RMB. All effects are statistically significant (bootstrap $95\%$ CIs exclude zero).}
\end{minipage}
\end{table}

\begin{figure}[htbp]
\centering
\caption{Placebo test: density at a pseudo threshold ($600$~RMB), URRBMI 2012 (baseline specification).}
\includegraphics[width=0.6\linewidth]{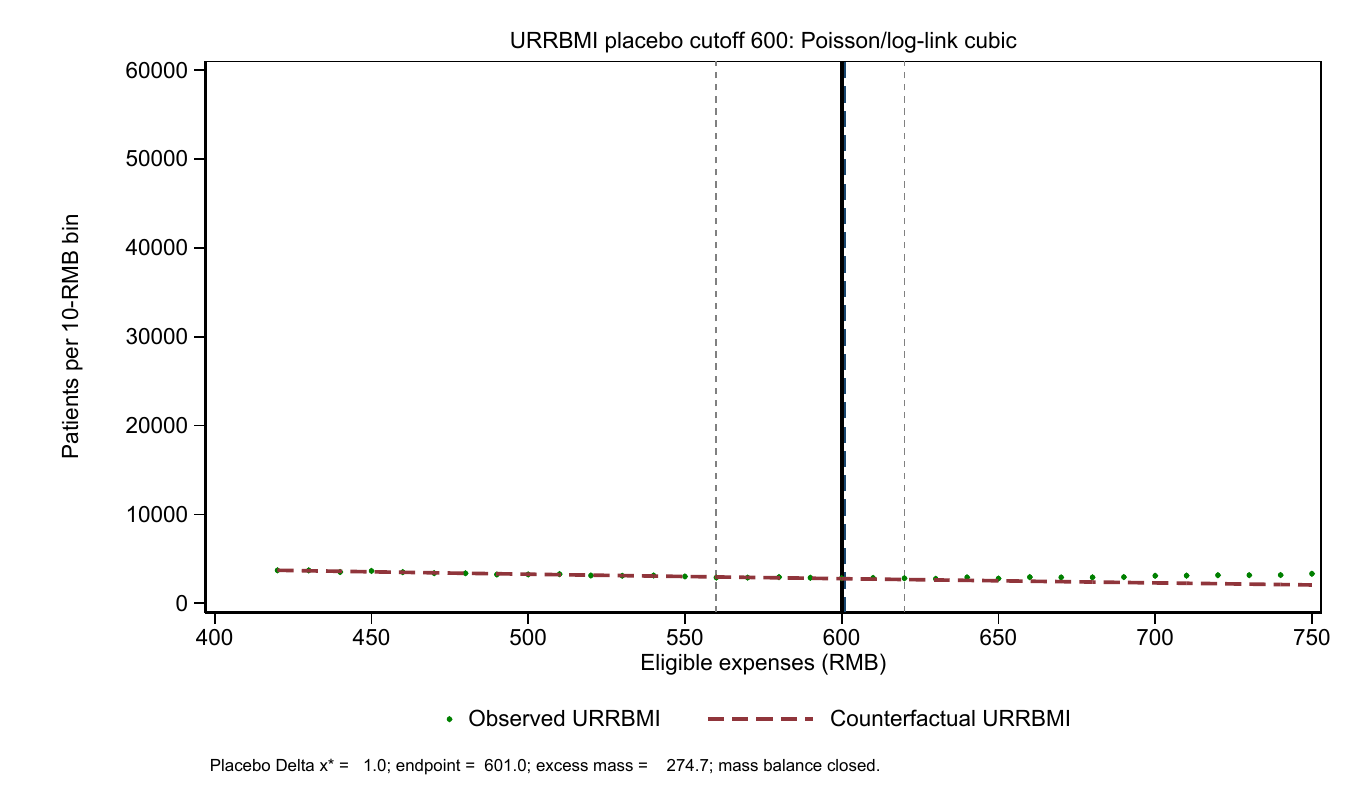}
\begin{minipage}{16cm}
\footnotesize{Note: The procedure of Section~\ref{sec:app-density} is applied at $600$~RMB---the URRBMI cap in 2011 but not in 2012, so non-binding in 2012. The solid line is the observed URRBMI density and the dashed line the design-based counterfactual; the solid vertical marks the placebo threshold and the inner dashed verticals the diffuse window $[560,620]$. The two curves coincide and the implied $\widehat{\Delta x}^{*}$ is $1.0$~RMB (versus $174.3$ at the true $800$~RMB threshold), so the method detects no bunching where no kink exists.}
\end{minipage}
\label{fig:app-placebo}
\end{figure}

\textbf{Placebo threshold.} A genuine policy kink should leave a footprint only at the policy threshold. URRBMI's cap was $600$~RMB in 2011 but moved to $800$~RMB in 2012, so in 2012 the value $600$ is no longer binding and serves as a natural placebo: applying the entire density procedure of Section~\ref{sec:app-density} at $600$~RMB in 2012 should detect no bunching. Figure~\ref{fig:app-placebo} confirms this: the observed and design-based counterfactual densities coincide through $600$~RMB, with no spike and no offsetting deficit. The implied marginal response is $\widehat{\Delta x}^{*}=1.0$~RMB at the baseline specification (range $[1.0,1.4]$ across the six specifications), far below the $\widehat{\Delta x}^{*}\approx174$~RMB found at the true $800$~RMB threshold. Bunching therefore tracks the \emph{current} policy threshold: it is absent at the prior cap once it stops binding, just as it is absent at $800$~RMB in 2011 before the cap moved there (Section~\ref{sec:app-bunching}). The same null holds at a never-binding value of $650$~RMB ($\widehat{\Delta x}^{*}=5.4$~RMB), so the falsification is not specific to the chosen threshold.

\begin{figure}[htbp]
\centering
\caption{Placebo test: outcome functions at a pseudo threshold ($600$~RMB), URRBMI 2012 (baseline specification). Left: total outpatient visits. Right: average cost per visit.}
\includegraphics[width=0.49\linewidth]{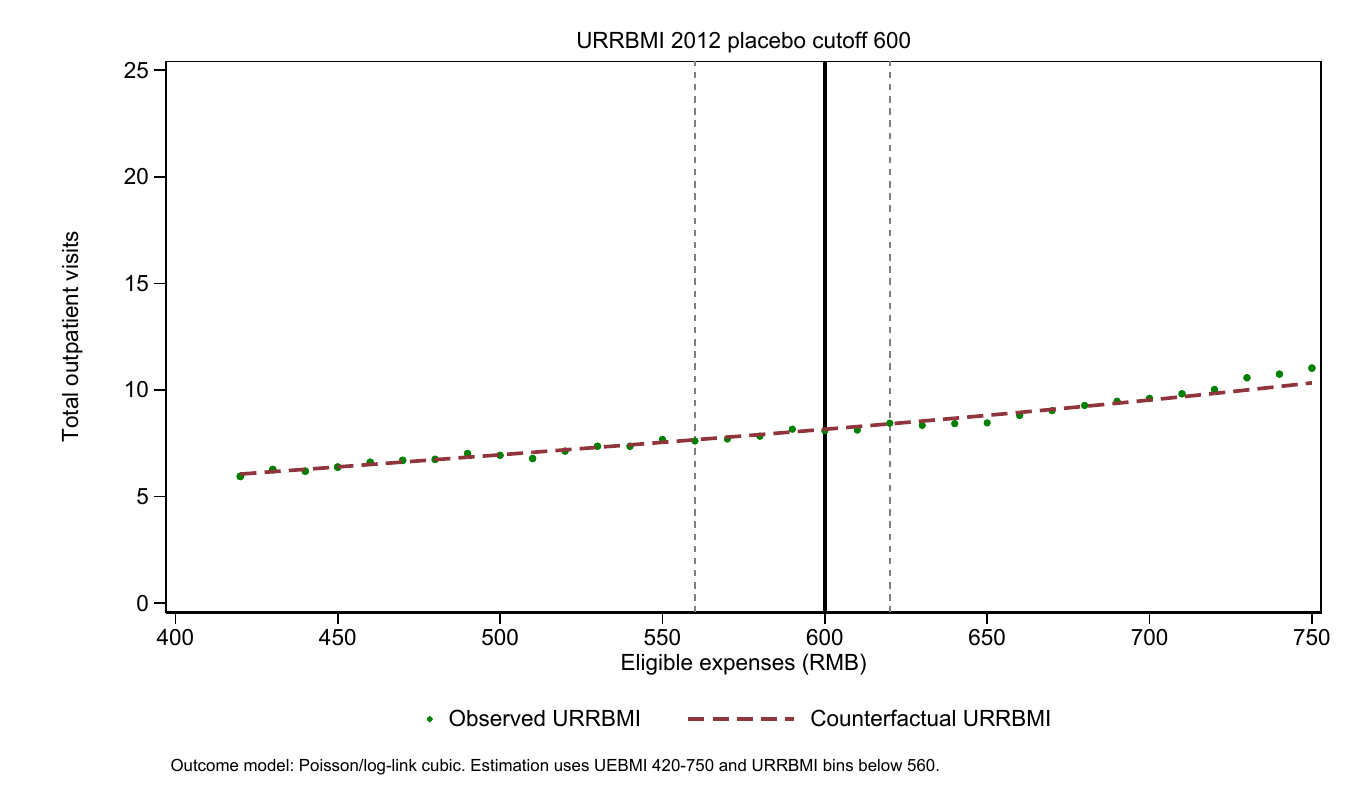}
\includegraphics[width=0.49\linewidth]{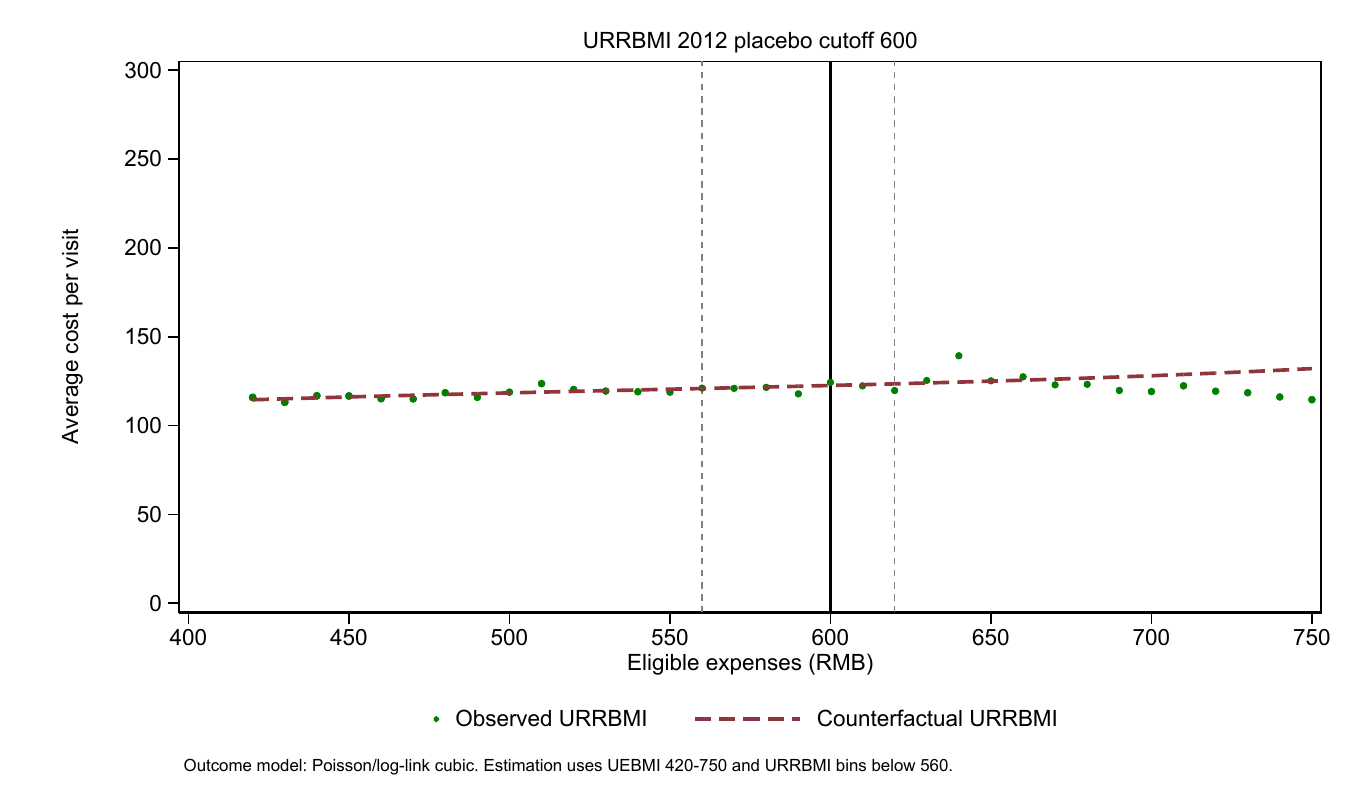}
\begin{minipage}{16cm}
\footnotesize{Note: The counterfactual-outcome construction of Section~\ref{sec:app-outcomes} is applied at the $600$~RMB pseudo-threshold, non-binding in 2012. Green markers are observed bin-level means for URRBMI (left: outpatient visits; right: cost per visit, RMB); the dashed line is the design-based counterfactual outcome function. The solid vertical marks the pseudo-threshold and the inner dashed verticals the diffuse window $[560,620]$. In both panels the observed and counterfactual profiles coincide above $600$~RMB, so the design recovers no treatment effect where no kink exists (shifter effects $\widehat\tau_S=0.16$ visits and $-3.1$~RMB, neither significant).}
\end{minipage}
\label{fig:app-placebo-outcome}
\end{figure}

The outcome side passes the same test. We apply the counterfactual-outcome construction of Section~\ref{sec:app-outcomes} at the $600$~RMB pseudo-threshold: with no kink, the observed and design-based counterfactual outcome functions should coincide, leaving no effect to detect. Figure~\ref{fig:app-placebo-outcome} confirms this for both primary outcomes---the observed and counterfactual profiles lie on top of one another above $600$~RMB. The implied shifter effect is statistically zero, $\widehat\tau_S=0.16$ visits ($95\%$ CI $[-0.34,0.60]$) for total outpatient visits and $-3.1$~RMB ($[-10.5,4.7]$) for cost per visit, against $\widehat\tau_S=-8.92$ visits and $66.8$~RMB at the true cap; the remaining outcomes behave likewise.

The placebo test at the lapsed threshold does double duty as a check on two extensions developed in Section~\ref{sec:extensions}. First, the absence of residual bunching at $600$~RMB in 2012 means that no enrollees remain anchored at the previous kink---there are no \emph{stayers}---so the frictionless benchmark underlying the mass-balance condition is appropriate here, and the friction correction of Section~\ref{sec:diffuse-frictions} is not needed. 
Second, $600$ and $650$~RMB are salient round numbers, yet the procedure detects essentially no excess mass at either value; reference-point bunching addressed in Section~\ref{sec:rounding-reference-points} is therefore negligible in medical context, which supports interpreting the excess mass at $800$~RMB as a policy response rather than a rounding artifact.


\subsection{Heterogeneity}
\label{sec:app-heterogeneity}
The design-based estimator extends to subgroups: we re-run the entire pipeline---the counterfactual density and marginal response $\Delta x^*$, the counterfactual outcome function, and the treatment effects---separately within each cell, using the corresponding UEBMI subgroup as the placebo and the subgroup's own unaffected region below $x^*$ to fit the level-and-slope adjustments. Table~\ref{tab:app-heterogeneity} reports the estimates for the two primary outcomes.

The bunching response and the treatment effects are similar for men and women. The marginal response is $\widehat{\Delta x}^{*}=178$~RMB for men and $171$~RMB for women, both close to the pooled $174$~RMB. The visit reduction is comparable---shifters make about nine fewer outpatient visits in each group ($\widehat\tau_S=-9.1$ for men, $-8.9$ for women)---and the rise in cost per visit is likewise similar ($\widehat\tau_S=+63.8$ and $+67.9$~RMB). Because men have a higher counterfactual cost per visit to begin with, the proportional increase is marginally larger for women. Buncher effects track the pooled estimates in both groups.

Age groups differ more. The marginal bunching response is largest for working-age enrollees ($\widehat{\Delta x}^{*}=194$~RMB, ages $16$--$54$) and children ($189$, ages $\le 15$), and smallest for the elderly ($161$, ages $\ge 55$). The buncher visit reduction rises steadily with age---from $\widehat\tau_B=-0.8$ visits for children to $-2.5$ for working-age enrollees and $-3.2$ for the elderly---while the shifter visit reduction is more stable across groups ($-7.5$ to $-8.5$). Children's bunchers barely change their cost per visit ($\widehat\tau_B=+10$~RMB), whereas working-age and elderly bunchers raise it sharply ($+49$ and $+41$~RMB); the shifter cost-per-visit increase is large in every group ($+46$ to $+53$~RMB). The muted buncher response among children is consistent with their outpatient care being less discretionary at the margin.


\begin{table}[htbp]
\centering
\caption{Heterogeneity in the design-based estimates, URRBMI 2012 (baseline specification).}
\label{tab:app-heterogeneity}
\small
\begin{tabular}{l c cc cc}
\toprule
 & & \multicolumn{2}{c}{Total visits (\texttt{novisit})} & \multicolumn{2}{c}{Cost/visit (\texttt{zfy\_avg})} \\
\cmidrule(lr){3-4}\cmidrule(lr){5-6}
Subgroup & $\widehat{\Delta x}^{*}$ & $\widehat\tau_B$ & $\widehat\tau_S$ & $\widehat\tau_B$ & $\widehat\tau_S$ \\
\midrule
Full sample & $174.3$ & $-2.58$ & $-8.92$ & $36.3$ & $66.8$ \\
            & $(3.1)$ & $(0.14)$ & $(0.21)$ & $(2.0)$ & $(3.1)$ \\
\addlinespace
\multicolumn{6}{l}{\textit{By gender}} \\
\quad Male   & $178.3$ & $-2.59$ & $-9.11$ & $33.5$ & $63.8$ \\
             & $(5.3)$ & $(0.24)$ & $(0.35)$ & $(3.4)$ & $(5.0)$ \\
\quad Female & $171.4$ & $-2.66$ & $-8.86$ & $38.2$ & $67.9$ \\
             & $(4.5)$ & $(0.19)$ & $(0.28)$ & $(2.4)$ & $(3.8)$ \\
\addlinespace
\multicolumn{6}{l}{\textit{By age group}} \\
\quad Aged $\le 15$  & $189.4$ & $-0.82$ & $-7.84$ & $10.2$ & $51.6$ \\
                     & $(7.3)$ & $(0.27)$ & $(0.42)$ & $(5.1)$ & $(7.2)$ \\
\quad Aged 16--54    & $194.2$ & $-2.51$ & $-7.51$ & $49.1$ & $45.8$ \\
                     & $(7.8)$ & $(0.25)$ & $(0.40)$ & $(4.8)$ & $(6.9)$ \\
\quad Aged $\ge 55$  & $161.0$ & $-3.20$ & $-8.47$ & $40.5$ & $52.8$ \\
                     & $(3.8)$ & $(0.22)$ & $(0.34)$ & $(2.7)$ & $(4.4)$ \\
\bottomrule
\end{tabular}
\begin{minipage}{15cm}
\vskip 4pt
\footnotesize{Note: Each subgroup is re-estimated with the full design-based pipeline (baseline Poisson/log-link cubic density, estimation window $[420,1280]$, diffuse window $[760,820]$, $10$-RMB bins, $B=300$), using the corresponding UEBMI subgroup as the placebo; bootstrap standard errors in parentheses. $\widehat{\Delta x}^{*}$ is the marginal bunching response (RMB). $\widehat\tau_B$ and $\widehat\tau_S$ are the average effects on bunchers and shifters; total visits are counts and cost per visit is in RMB. Heterogeneity by location (urban \texttt{URBMI} vs.\ rural \texttt{NRCMS}) will be added.}
\end{minipage}
\end{table}

Note the behavior of $\widehat{\Delta x}^{*}$ across cells is itself the diagnostic suggested in Section~\ref{sec:heterogeneity-covariates}: its stability by gender and its more pronounced variation by age indicate that responsiveness differs mainly across age groups, so pooled estimates should be read as approximations whose accuracy is governed by that variation.


\section{Extensions and Sensitivity}
\label{sec:extensions}

The baseline analysis identifies average effects for affected populations by combining design-assisted counterfactuals with mass-preserving assignment intervals. The density module recovers \(h_F^{cf}\) and the marginal response \(\Delta x^*\); the outcome module recovers \(m_F^{cf}\); and the treatment-effect formulas compare observed bunchers and shifters with the corresponding counterfactual intervals. 

The extensions below clarify what changes when one component of this procedure is modified. Response heterogeneity, stayers, rounding, and alternative counterfactual policies change either the affected population or the relevant counterfactual policy. Structural relocation adds assumptions that deliver diagnostics and pointwise objects beyond the baseline. 


\subsection{Heterogeneity in Assignment Responses}
\label{sec:heterogeneity-covariates}
The analysis so far has treated response elasticity as common to all agents: under Assumption~\ref{assume:monotone}, the scalar type $\eta$ is the only source of heterogeneity, so agents differ in where they locate but not in how strongly they respond to the rate change. In practice, agents may also differ in their response elasticity, apart from innate heterogeneity in ability. For example, empirical literature documents that female labor supply responds more elastically to tax changes than male labor supply. 

Suppose, then, that the response elasticity $e$ varies across groups defined by predetermined covariates $W$, such as gender or age, with $e$ constant within each group, while $\eta$ varies within groups and every group spans the full range of $X^{cf}$. The effective type is two-dimensional, $(\eta, e(W))$, and Assumption~\ref{assume:monotone} fails in the pooled population: at any counterfactual value $x$, agents from different groups coexist---the heterogeneity is \emph{within} bins of $X^{cf}$---and agents with the same $X^{cf}$ no longer respond identically. This failure does not contaminate the counterfactual objects: the estimation of $h_F^{cf}$ and $m_F^{cf}$ relies on the placebo group and the focal group's unaffected region, not on Assumption~\ref{assume:monotone}, and remains valid. What fails is the assignment of observed agents to their counterfactual locations. The counterfactual region for bunchers is now group-specific, $(x^*,\,x^*+\Delta x^*(w)]$, so the pooled mass-balance equation delivers only an equivalent-mass interval $(x^*,\,x^*+\Delta x^*]$ that matches no group's actual boundary. The buncher effect $\tau_B$ is then biased because bunchers are not compared with their own counterfactuals. 
The shifter effect $\tau_S$ suffers the same mismatch: an observed shifter window originates from group-specific counterfactual bands $(x^*+\Delta x^*(w),\,x^{cf,out}(w)]$, which the pooled equal-mass condition replaces with a single interval of the correct mass but the wrong support and composition.

The remedy is to condition on $W$. Within a cell, $e$ is constant and $\eta$ is again the only dimension of heterogeneity, so Assumption~\ref{assume:monotone}---and with it rank invariance (Corollary~\ref{cor:rank-invariance})---holds. Estimation repeats the baseline pipeline within each cell: the cell's placebo counterpart and its own unaffected region deliver $h_F^{cf}(x\mid w)$ and $m_F^{cf}(x,w)$; the cell's excess mass and mass-balance equation deliver $\Delta x^*(w)$; and the cell-level effects $\tau_B(w)$ and $\tau_S(w)$ follow as in the baseline analysis. The aggregate effects then average the cell-level effects with weights proportional to each cell's \emph{affected} mass,
\begin{equation}
    \tau_B=\frac{\int \tau_B(w)\,B_F(w)\,f_F(w)\,dw}{\int B_F(w)\,f_F(w)\,dw},
    \qquad
    B_F(w)=\int_{x^*}^{x^*+\Delta x^*(w)} h_F^{cf}(x\mid w)\,dx,
    \label{eq:het-tauB-agg}
\end{equation}
and analogously for $\tau_S$ over the cell-specific shifter windows. Buncher mass is the appropriate weight: more responsive groups bunch from wider counterfactual bands and contribute disproportionately to the affected population. When $W$ is discrete and each cell is estimated flexibly, this conditional procedure \emph{is} subgroup estimation. When cells are small, one can pool the estimation of nuisance objects across cells---restricting the focal--placebo adjustment coefficients to be common, 
while keeping the identification-critical margins, $\Delta x^*(w)$ fully flexible.  
 The functional-form restrictions fall on nuisance functions rather than on the response margin. 

Note $W$ absorbs only observable heterogeneity in responsiveness: the maintained condition weakens from homogeneity to the absence of $W$-residual response heterogeneity---more defensible, but an assumption rather than a testable fact. 
 A useful diagnostic is to estimate $\Delta x^*(w)$ across cells and finer splits: systematic variation signals that pooling misassigns the affected population, while stability is reassuring about observables though silent on unobserved responsiveness. 


Heterogeneity in how the assignment $X$ and the transfer $T$ affect the outcome $Y$ is, by contrast, left unrestricted. Because the estimands compare the \emph{same} agents' treated and counterfactual outcomes, the total effect may vary freely with $X^{cf}$, $W$, and unobservables. 

\subsection{Stayers / Optimization Frictions}
\label{sec:diffuse-frictions}
Optimization frictions raise a separate issue. The baseline model treats all agents in the affected counterfactual region as responsive. Following the bunching literature, let a scalar share \(\pi\in[0,1)\) be unresponsive. These agents keep their counterfactual assignment, while the remaining share \(1-\pi\) responds to the kink. When the kink changes over time, bunching at a previous kink can be used to calibrate \(\pi\). 

Since only the responsive share bunches, excess mass at the current kink satisfies:
\begin{equation}
    B_F   =  \int_{x^*}^{x^*+\Delta x^*_{\pi}}  (1-\pi)  h_F^{cf}(u)\,du, 
    \label{eq:friction-buncher-mass}
\end{equation}
where \(\Delta x^*_{\pi}\) denote the marginal response under friction. 
Thus, for a fixed counterfactual density, allowing for stayers widens the response interval relative to the frictionless benchmark \(\pi=0\).


The observed density above the kink $h_F^{kp}(x), \quad x>x^*$ consists of responsive shifters who reduce their assignment value to $X^{kp}=x$ and unresponsive stayers who stay at their counterfactual value with $X^{kp}=X^{cf}=x$. Thus, the density of responsive shifters under the kinked policy is:\footnote{Suppose, for this sensitivity exercise, that responsive shifters follow a known monotone relocation map \(X^{kp}=R(X^{cf})\). Then for the continuous density above the cutoff,
\begin{equation*}
    h_F^{kp}(x) = (1-\pi) h_F^{cf}\!\left(R^{-1}(x)\right)  \left|\frac{dR^{-1}(x)}{dx}\right|     +   \pi h_F^{cf}(x).
    \label{eq:friction-hkp-decomp}
\end{equation*}
The first term is the observed density of responsive shifters; the second term is the density of unresponsive stayers.   That is, $ h_F^{kp}(x)-\pi h_F^{cf}(x)=  (1-\pi)  h_F^{cf}\!\left(R^{-1}(x)\right)  \left|\frac{dR^{-1}(x)}{dx}\right|$. Note \(R\) is needed only if the researcher wants to interpret this density through a pointwise map.}
\begin{equation}
    h_{F,R}^{kp}(x)
    \equiv
    h_F^{kp}(x)-\pi h_F^{cf}(x),
    \label{eq:friction-responsive-density}
\end{equation}

For the interval-average shifter estimand, the responsive equal-mass endpoint \(x_{\pi}^{cf,out}\) solves
\begin{equation}
    \int_{x^*}^{x^{kp,out}}
    h_{F,R}^{kp}(z)\,dz
    =
    (1-\pi)
    \int_{x^*+\Delta x^*_{\pi}}^{x_{\pi}^{cf,out}}
    h_F^{cf}(u)\,du .
    \label{eq:friction-shifter-mass}
\end{equation}

The literature on bunching with frictions mainly focus on the assignment variable density, while outcome side is less explored.  
 If unresponsive and responsive agents from the same counterfactual assignment bin have different potential outcomes, then \(\pi\) changes not only the mass balance but also the composition of affected agents. Let \(m_{F, R}^{cf}(u)\) and \(m_{F, UR}^{cf}(u)\) denote counterfactual outcome means for responsive and unresponsive agents with counterfactual assignment \(u\). A transparent sensitivity parameterization is
\begin{equation}
    m_{F,UR}^{cf}(u)=\beta m_{F,R}^{cf}(u).
    \label{eq:friction-beta-cf}
\end{equation}
Since the design-assisted counterfactual outcome function averages the two latent types,
\begin{equation}
    m_F^{cf}(u)
    =
    (1-\pi)m_{F,R}^{cf}(u)+\pi m_{F,UR}^{cf}(u),
    \qquad
    m_{F,R}^{cf}(u)
    =
    \frac{m_F^{cf}(u)}
    {(1-\pi)+\pi\beta} .
    \label{eq:friction-responsive-mcf}
\end{equation}
The counterfactual mean for responsive bunchers is then obtained by replacing \(m_F^{cf}(u)\) with \(m_{F,R}^{cf}(u)\) and weighting by \((1-\pi)h_F^{cf}(u)\) over \((x^*,x^*+\Delta x^*_{\pi}]\). The treated mean for bunchers is unchanged. 

For shifters, the observed outcome mean above the cutoff $m^{kp}(x), \quad x>x^*$ can also mix responsive shifters and unresponsive stayers. The treated mean outome for responsive shifters is
\begin{equation}
    m_{F,R}^{kp}(x)  =
    \frac{m_F^{kp}(x)h_F^{kp}(x)- m_{F,UR}^{cf}\cdot \pi h_F^{cf}(x)}{h_{F,R}^{kp}(x)}=\frac{m_F^{kp}(x)h_F^{kp}(x)- \frac{\pi\beta}{(1-\pi)+\pi \beta} m_F^{cf}(x)  h_F^{cf}(x) }{h_F^{kp}(x)-\pi h_F^{cf}(x)}
    \label{eq:friction-responsive-mkp}
\end{equation}
The responsive-shifter treatment effect replaces the baseline treated mean with the average of \(m_{F,R}^{kp}(x)\) weighted by \(h_{F,R}^{kp}(x)\) over \((x^*,x^{kp,out}]\), and replaces the baseline counterfactual mean with the average of \(m_{F,R}^{cf}(u)\) weighted by \((1-\pi)h_F^{cf}(u)\) over \((x^*+\Delta x^*_{\pi},x_{\pi}^{cf,out}]\).


\subsection{Relabeling and misreporting}
\label{sec:relabeling-reference-points}



Suppose the observed assignment variable is a reported value \(X^{rp}\), while the unobserved real value is \(X^{rl}\). 
The resulting treatment effect should then be interpreted as the reduced-form effect of the kinked policy, combining any real response with reporting or relabeling responses. 

Denote the degree of relabeling as $\delta \equiv (X^{rp}-X^{rl})/X^{rp}$.
If the researcher wants to further isolate the real-response channel, additional information is needed, such as third-party records on  misreporting, or a structural model on misreporting.  For example, \cite{chen2021notching} shows that under weak assumptions, agents' optimal degree of relabeling depends on the parameter in relabeling cost function $c$ and the marginal tax rate $\tilde{t}$ linked to tax saving. That is, $\delta = f(\frac{\tilde{t}}{c})$. Further, since agents can counter-balance the increase in tax rate via altering their optimal degree of relabeling,  the  marginal response in reported assignment value, $\Delta x^{rp,*}$, is typically smaller than what would be observed if relabeling is \textit{not} allowed; that is, $\Delta x^{rp,*}<\Delta x^*$. Given two bunching moments, one can recover both  the misreporting cost paramter (thus the degree of relabeling) and the response elasticity.


\subsection{Rounding, and Reference-Point Bunching}
\label{sec:rounding-reference-points}

 Agents often bunch at round values for reasons unrelated to the policy. When $x^*$ coincides with such a value, the excess mass combines policy-induced bunching with reference-point bunching that would have occurred even absent the kink, and attributing all of it to the policy overstates $\Delta x^*$. Treated-group-only bunching estimators address this by adding round-number indicators to a polynomial counterfactual \citep{chetty2011adjustment}; the design-assisted approach can do the same while replacing the polynomial with the placebo-disciplined shape.

The placebo group is not subject to the focal kink but shares the same rounding environment, so it reveals the reference-point pattern in the local window. Whether an explicit correction is needed depends on how the placebo density enters. If the observed placebo density $h_P(x)$ is used directly, rounding is already present and no adjustment is required. If instead $h_P$ is approximated by a smooth functional form,
---as in the baseline, which fits a low-order density model---the smooth fit averages over the rounding-induced bunching, so the fitted placebo density, and the focal counterfactual that inherits it, would be too smooth and 
the excess mass at $x^*$ overstated. The remedy is to include reference-point terms in the fitted model so that the fitted densities retain the rounding-induced bunching.

Concretely, one estimates a joint density model for $g\in\{F,P\}$, using the placebo group over the full window $\mathcal W$ and the focal group over its unaffected region $\mathcal U$:
\begin{equation}
    \ell_h\{h_g(x)\}
    =
    s_h(x)'\beta_h
    +
    R_h(x)'\gamma_h
    +
    \mathbf 1_{\{g=F\}}\,q_h(x)'\delta_h,
    \qquad x\in\mathcal W \text{ for } g=P,\quad x\in\mathcal U \text{ for } g=F,
    \label{eq:rounding-density}
\end{equation}
where $s_h(x)$ is the smooth basis (polynomial, bilog-polynomial, or spline), $R_h(x)$ is a vector of reference-point indicators for bins at multiples of $50$, $100$, and $200$, and $\ell_h$ is the link function. Because a value such as $x^*$ can be a multiple of $50$, $100$, and $200$ at once, the indicators are nested and its bunching is the sum of the applicable amplitudes. With a log link the reference-point terms enter multiplicatively (heaping proportional to the local density); with the identity link they are additive. The focal counterfactual density is
\begin{equation}
    \widehat h_F^{cf}(x)
    =
    \ell_h^{-1}\!\left\{
    s_h(x)'\widehat\beta_h
    +
    R_h(x)'\widehat\gamma_h
    +
    q_h(x)'\widehat\delta_h
    \right\},
    \label{eq:rounding-hcf}
\end{equation}
so the counterfactual used to compute excess bunching already carries the normal rounding component: the excess mass at $x^*$ is the observed focal density minus this rounded counterfactual. 

The reference-point amplitude $\gamma_h$ is identified entirely from round bins where no policy operates. Equation~\eqref{eq:rounding-density} already enforces this: the placebo enters over all of $\mathcal W$---including its bin at $x^*$, where it faces no kink and so records only rounding---while the focal group enters only through $\mathcal U$, so its own, policy-contaminated bin at $x^*$ is excluded. The counterfactual rounding at $x^*$ is therefore an out-of-sample prediction. Two cases arise, distinguished by the reference-point basis:
\begin{enumerate}
    \item \emph{Generic round number.} If $x^*$ shares its round-number class with other, non-policy values in the window---for example, a multiple of $200$ alongside $600$, $1000$, and $1200$---a class indicator suffices and its amplitude is identified from those values. When the focal group's unaffected region contains enough same-class round numbers, this can be done from the focal group alone, so no focal--placebo rounding equality is required.
    \item \emph{Special round number.} If $x^*$ is uniquely salient and cannot be approximated from other round values, its rounding-induced bunching needs a dedicated indicator; because the focal bin at $x^*$ is excluded, that indicator is identified only from the placebo's bin at $x^*$.
This requires the focal and placebo groups to share the reference-point amplitude at $x^*$, after the low-dimensional adjustment. This assumption is likely to hold, when the focal and placebo groups share similar reference-point amplitude at locations other than $x^*$. 
\end{enumerate}


The same device applies to the outcome equation when conditional outcome means display reference-point artifacts. We can augment the fitted outcome model with analogous terms,
\begin{equation}
    \ell_y\{m_g(x)\}
    =
    s_y(x)'\beta_y
    +
    R_y(x)'\gamma_y
    +
    \mathbf 1\{g=F\}\,q_y(x)'\delta_y,
    \label{eq:rounding-outcome}
\end{equation}
and predicts $m_F^{cf}(x)$ using the focal adjustment and the common reference-point component. 

\subsection{Alternative counterfactual policies}
\label{sec:alternative-counterfactual-policy}

The baseline estimand compares the kinked policy with a linear counterfactual schedule that applies the below-threshold marginal rate \(t\) throughout the local window: $T^{cf}(x)=tx.$
It evaluates what would happen if the higher marginal rate above the threshold were removed.

Researcher may want to compare the kinked policy with a linear counterfactual schedule that applies the above-threshold marginal rate \(t+\Delta t\) throughout the local window:
\vspace{-5pt}
\begin{equation} \label{policy:act}
    T^{cf,H}(x)= (t+\Delta t)x. 
\end{equation}
Let \(X^{cf,H}\) and \(Y^{cf,H}\) denote the assignment and outcome under the high-rate linear counterfactual. The relevant counterfactual density becomes \(h_F^{cf,H}(x)\), the counterfactual outcome function becomes
\(
    m_F^{cf,H}(x)=\mathbb E[Y^{cf,H}\mid X^{cf,H}=x],
\)
and the treatment effect changes accordingly.

The placebo group faces the scheme $T^{cf, H}(x)$. 
For the focal group under the kinked scheme $T^{kp}(x)=t x+ \Delta t  x I_{\{x>x^*\}} - \Delta t x^* I_{\{x>x^*\}}$, we have
\begin{itemize}
\item  Agents above the kink $x^*$ pay $(t+\Delta t)x-\Delta t x^*= T^{cf,H}(x)-\Delta t x^*$. They do not face changes in marginal incentives (i.e., always at $t+\Delta t$), hence, their assignment variable remains the same, with $X^{kp}=X^{cf, H}>x^*$. However, the lump-sum transfer, $ -x^*\Delta t$, may lead to changes in their outcome, where $E[Y^{kp}|X^{kp}]=E[Y^{cf, H}|X^{kp}]+\tau(-x^*\Delta t), \forall X^{kp}>x^*$

\item  Agents with $X^{cf,H}<x^*-\widetilde{\Delta x^*}$ are shifters, who face a reduction in marginal incentive from $(t+\Delta t)$ to $t$ and increase their assignment value, with $X^{cf,H}<X^{kp}<x^*$.  Their outcome values change accordingly, with $\mathbb E [y^{kp}|x^{cf,H}]\neq E [y^{cf,H}|x^{cf,H}], \forall x^{cf,H}<x^*-\widetilde{\Delta x^*}$.

\item Agents with $x^{cf,H}\in [x^*-\widetilde{\Delta x^*}, x^*]$ bunch at the kink $x^*$. 

\end{itemize}

The baseline framework applies to this alternative counterfactual after redefining the potential assignments and potential outcomes relative to \(T^{cf,H}\) and with modifications to accomodate the lump-sum transfer effect.  First, recover the relevant counterfactual assignment distribution, $h^{cf, H}(x)$, using the placebo group's density, $h_P(x), \quad \mathbb W$, and the focal group's density above the kink, $h^{kp}_F(x), \quad x>x^*$ and based on Assumption 2. 
Second, estimate the marginal bunching response $\widetilde{\Delta x^*}$ using the mass-balance equation for bunchers: $B_F=\int_{x^*-\widetilde{\Delta x^*}}^{x^*} h^{cf, H}(x) dx $. Similarly, find the corresponding counterfactual ending point for shifters using the mass-balance equation, $\sum_{x^{kp,in}}^{x^*} h^{kp}(x) dx= \int_{x^{cf,in}}^{x^*-\widetilde{\Delta x^*}} h^{cf,H}(x) dx$.
Third, assuming that the lump-sum transfer effect, $\tau(-\Delta t x^*)$, is a constant, estimate the counterfactual outcome function  together with the lump-sum transfer effect,$\mathbb E [y^{cf,H}|x^{cf,H}]+\tau(-\Delta t x^*)$ using the placebo group's outcome function, $E[Y_P | X_P=x], \quad \mathbb W$, and the focal group's outcome function above the kink, $E[Y^{kp}_F|X^{kp}_F=x], \quad x>x^*$ and based on Assumption 4. With assumptions on how $\Delta_X$ and $\Delta_T$ influences outcome $Y$, we can separate the effect of lump-sum transfer  $\tau(-\Delta t x^*)$, as it is also in monetary-term and thus shares the same coefficient as $\Delta_T$.
Fourth, estimate the average treatment effect on bunchers and shifters similar to the baseline, by comparing their average treated outcome to their counterfactual. 


This extension clarifies that the causal estimand is policy-counterfactual specific. The same observed kinked policy can have different treatment effects depending on whether it is compared with a low-rate linear schedule, a high-rate linear schedule, or another feasible reform.

\subsection{Structural Relocation: Diagnostics and Pointwise Effects}
\label{sec:structural-models}

The main design-assisted estimator identifies aggregate effects for shifters over an interval. 
A structural response model can deliver more. Suppose the economic environment implies a monotone map from observed  assignments for shifters under the kink to their latent counterfactual assignments,
\begin{equation}
    X^{kp}=R_\theta(X^{cf}),
    \qquad X^{cf}>x^*+\Delta x^*, \text{ or }  X^{kp}>x^*,
    \label{eq:general-relocation-map}
\end{equation}
where \(R_\theta(\cdot)\) is strictly increasing on the shifter region, and satisfies \(R_\theta(x^*+\Delta x^*)=x^*\). 
The parameter \(\theta\) is pinned down by the same density and mass-balance objects used above and the model. 
This is stronger than the baseline: it assigns each observed shifter to a point in the counterfactual assignment distribution. 

The familiar example is the Saez-style constant-elasticity model. If the assignment choice under a linear marginal rate \(\tilde t\) satisfies\( x(\eta,\tilde t)=\eta(1-\tilde t)^e,\)
then shifters facing the higher marginal rate above the kink satisfy
\begin{equation}
    X^{kp}  =  X^{cf}\frac{(1-t-\Delta t)^e}{(1-t)^e}  =    X^{cf}\frac{x^*}{x^*+\Delta x^*}.   \label{eq:structural-proportional-map}
\end{equation}
The last equality uses the marginal buncher condition: the agent who is indifferent at the kink would have chosen \(x^*+\Delta x^*\) under the counterfactual linear policy and \(x^*\) under the higher marginal rate.
Other models may imply different maps, e.g., a proportional map in an underlying index that is then transformed into the observed assignment variable,\footnote{In Lu et al. (2026), the model implies that, faced with firm's average wage deduction limit, firms adjust their skilled-unskilled labor ratio following the proportional mapping, with $\frac{H^{cf}}{L^{cf}}=\frac{H^{kp}}{L^{kp}}c$, where $c$ is a constant.
Given that firm-level average wage $x_i=\frac{w_H H^{cf}_i+w_L L^{cf}_i}{H^{cf}_i+ L^{cf}_i}=(w_H \frac{H^{cf}_i}{L^{cf}_i}+w_L ) / (\frac{H^{cf}_i}{L^{cf}_i}+ 1)$, where $w_H, w_L$ are scalars, we can still derive the mapping $x_i^{kp}=R_{\theta}(x_i^{cf})$, which is a known function of the proportional mapping.}
 a covariate-adjusted map \(R_\theta(x,w)\).\footnote{It is justified by $\max_x x -t x -\frac{\eta}{1+1/e}(\frac{x-\gamma W}{\eta})^{1+1/e}$, where $x$ is running variable, $\eta$ denotes individual heterogeneity in ability, $e$ is a scalar, and \(W\) is predetermined covariate. The first-order condition yields $x=\eta (1-e)^e+\gamma W$, which implies $x^{cf}-\gamma W = (x^{kp}-\gamma W)\frac{x^*+\Delta x^*}{x^*}$.} 

The first use of \(R_\theta\) is diagnostic. The map transforms the observed shifter density into an implied counterfactual density:
\begin{equation}
    \widetilde h^{cf}_F(u)  =  h^{kp}_F \!\left(R_\theta(u)\right)
    \left|   \frac{dR_\theta(u)}{du}  \right|,
    \qquad u>x^*+\Delta x^* .
    \label{eq:structural-relocated-density}
\end{equation}
This relocated density can be compared with the design-assisted \(\widehat h_F^{cf}(u)\) on the shifting region. Agreement indicates that the structural response model and the placebo-disciplined density restriction imply similar counterfactual densities. Disagreement 
may reflect misspecification of the relocation map, misspecification of the design-assisted density, or finite-sample instability. Importantly, the diagnostic does not resolve the single-kink non-identification problem by itself. The collapsed buncher interval \((x^*,x^*+\Delta x^*]\) remains unobserved, so its density shape must still come from the design-assisted restriction or explicit functional form restriction.

The second use of \(R_\theta\) is to define conidtional average treatment effect on shifters. Relocation gives an auxiliary treated outcome function indexed by latent counterfactual assignment:
\begin{equation}
    m_{F}^r(u)
    =
    \mathbb E\!\left[
        Y^{kp}\mid R^{-1}_\theta(X^{kp})=u,F
    \right],
    \qquad u>x^*+\Delta x^* .
    \label{eq:structural-treated-outcome}
\end{equation}
Combining this object with the design-assisted counterfactual outcome function yields
\begin{equation}
    \tau_{F}(u)
    =
    m_{F}^r(u)-m_F^{cf}(u),
    \qquad u>x^*+\Delta x^* .
    \label{eq:structural-pointwise-effect}
\end{equation}
Equation~\eqref{eq:structural-pointwise-effect} asks for the conditional average treatment effect for shifters at each counterfactual assignment value. This is a stronger estimand
than the baseline, which averages treated outcomes over an observedwindow and compares them with the equal-mass counterfactual interval.\footnote{The baseline does not require a point-wise mapping between observed and counterfactual locations of each shifter.}

Given the relocation map, one can also formulate a channel decomposition. For a counterfactul assignment value $X^{cf}=u$, its correspinding observed value is \(R_\theta(u)\equiv X^{kp}(u)\). Define
\( \Delta_X(u)=X^{kp}(u)-u,
    \quad
    \Delta_T(u)=T^{kp}\!\left(X^{kp}(u)\right)-T^{cf}(u).
\)
A local additive channel restriction writes
\begin{equation}
    \tau_{F}(u)
    =
    \mu(u)\Delta_X(u)
    +
    \lambda(u)\Delta_T(u), \quad u>x^*+\Delta x^*.
    \label{eq:channel-decomposition}
\end{equation}
 The total pointwise effect \(\tau_{F}(u)\) is identified only after imposing the relocation map and estimating \(m_F^{cf}\). The decomposition into assignment and transfer channels requires still more structure: either additional variation that moves \(\Delta_X\) and \(\Delta_T\) separately, or restrictions on \(\mu(u)\) and \(\lambda(u)\).

Under the proportional Saez map in Equation~\eqref{eq:structural-proportional-map}, \(X^{kp}(u)\) is linear in \(u\).  Because both the kinked policy and the counterfactual policy are linear in the shifter region, \(\Delta_X(u)\) and \(\Delta_T(u)\) are affine functions of \(u\).
 If, in addition, \(\mu(u)=\mu\) and \(\lambda(u)=\lambda\) are constant over the local shifter region, Equation~\eqref{eq:channel-decomposition} implies the familiar level-and-slope form
\begin{equation}
    \tau_{F}(u)=\beta_0+\beta_1u, \quad u>x^*+\Delta x^*.
    \label{eq:structural-level-slope}
\end{equation}
Thus the relocated shifter design resembles an RKD-style comparison indexed by \(X^{cf}\): the effect can be summarized by a level change and a slope change. But this is not assumption-free. If outcomes depend nonlinearly on the assignment or transfer, or if \(\mu(u)\) and \(\lambda(u)\) vary with \(u\), higher-order terms are required,
\begin{equation}
    \tau_{F}(u)
    =
    \beta_0+\beta_1u+\beta_2 u^2+\cdots, \quad u>x^*+\Delta x^*,
    \label{eq:structural-higher-order}
\end{equation}
or the pointwise function should be treated nonparametrically. The level-and-slope benchmark is a transparent structural sensitivity analysis: it links the method to familiar bunching and RKD language, while depending on the stronger relocation and channel assumptions.  

\section{Conclusion}
\label{sec:conclusion}

The bunching literature reads the distorted density of an assignment variable at a kink as a measure of behavioral response. This paper treats that response as the first stage of a causal design. When a kinked policy and the adjustment it induces jointly affect an outcome beyond the assignment variable, observed outcome profiles confound treatment effects with policy-induced movement in the assignment variable. The framework developed here resolves this by indexing agents by their latent counterfactual assignments, defining average total effects for the two affected populations---bunchers and shifters---and identifying the missing counterfactual objects with design variation: a placebo group disciplines the local shapes of the counterfactual density and outcome function, while the focal group's own unaffected observations pin down the permitted level and slope differences. Estimation is a transparent plug-in on binned data, with bootstrap inference. 

Two methodological messages deserve emphasis. First, a single kink under-identifies twice: it pins down neither the counterfactual assignment density nor, even after the assignment side is resolved, the counterfactual outcome function, since treated data reveal only the sum of that function and the treatment effect. We therefore purchase point identification with observable design variation rather than with functional form, and the restrictions bought this way can be inspected: they carry testable implications, which our application passes most sharply at the past threshold, where the full procedure detects neither bunching nor effects. Second, aggregate causal estimands require less behavioral structure than bunching practice often imposes. Rank invariance and mass balance suffice to pair observed bunchers and shifters with their counterfactual intervals; a pointwise relocation model is needed only for stronger objects, such as pointwise effect functions or channel decompositions.

The application illustrates what the outcome side adds. In China's URRBMI, enrollees bunch sharply at the annual reimbursement ceiling, and the excess bunching tracks the ceiling as it moves. 
 Above the cap, shifters cut outpatient visits by more than half relative to their counterfactual---and bunchers by about a fifth---while concentrating their remaining visits on more serious, higher-cost conditions, substituting hospital care for community health centers, and stopping care earlier in the year. Reimbursement ceilings are a widespread cost-containment device in outpatient coverage, and these estimates show that they 
lead to forgone care among precisely the enrollees who use medical care most. This is what matters for setting the level of the cap or the coinsurance rate above it.  

The design travels. Its requirements are the inputs of a standard bunching study, an outcome measured alongside the assignment variable, and credible placebo variation---agents subject to different thresholds, or the same program in a year when the threshold differed. Because policy thresholds are routinely revised and parallel schedules routinely coexist, such variation is common in tax, social insurance, and subsidy settings. The extensions show that departures from the baseline---heterogeneous responsiveness, optimization frictions, rounding and reference points, misreporting, alternative counterfactual schedules, structural relocation---each modify one module of the procedure while leaving the rest intact. The accompanying diagnostics indicate what applications should report: unaffected-region residuals, placebo thresholds, and stability across placebo groups and specifications.

Several caveats bound the results. The estimands are local: they describe bunchers and shifters in a window around the kink, not the full population, and they are total effects of the policy---decomposing them into assignment and transfer channels requires additional structure as discussed in the extension. Identification rests on parallel-shape restrictions that are testable but not guaranteed; where no credible placebo variation exists, the general framework supports partial-identification analysis rather than point estimates. 
Natural next steps include welfare analysis built on the estimated utilization responses. 
More broadly, the paper connects two strands of quasi-experimental work that have developed separately: regression kink designs estimate effects on outcomes but must rule out manipulation; bunching designs embrace manipulation but stop at the assignment variable. Causal inference under kink bunching occupies the intersection---turning manipulation from a threat to identification into its engine.

\bigskip 
\begingroup
\singlespacing
\bibliographystyle{plainnat}
\bibliography{acompat,reference}

\newif\ifabfull\abfulltrue
\begin{thebibliography}{22}
\providecommand{\natexlab}[1]{#1}
\providecommand{\url}[1]{\texttt{#1}}
\expandafter\ifx\csname urlstyle\endcsname\relax
  \providecommand{\doi}[1]{doi: #1}\else
  \providecommand{\doi}{doi: \begingroup \urlstyle{rm}\Url}\fi

\bibitem[Anagol et~al.(2024)Anagol, Davids, Lockwood, and
  Ramadorai]{anagol2024diffuse}
Santosh Anagol, Allan Davids, Benjamin~B Lockwood, and Tarun Ramadorai.
\newblock Diffuse bunching with frictions: Theory and estimation.
\newblock \emph{NBER Working Paper}, \penalty0 (w32597), 2024.

\bibitem[Bertanha et~al.(2022)Bertanha, Caetano, Jales, and
  Seegert]{bertanha2022bunching}
Marinho Bertanha, Carolina Caetano, Hugo Jales, and Nathan Seegert.
\newblock Bunching designs: A guide to practice.
\newblock 2022.

\bibitem[Bertanha et~al.(2023)Bertanha, McCallum, and
  Seegert]{bertanha2023better}
Marinho Bertanha, Andrew~H McCallum, and Nathan Seegert.
\newblock Better bunching, nicer notching.
\newblock \emph{Journal of Econometrics}, 237\penalty0 (2):\penalty0 105512,
  2023.

\bibitem[Best et~al.(2020)Best, Cloyne, Ilzetzki, and
  Kleven]{best2020estimating}
Michael~Carlos Best, James~S Cloyne, Ethan Ilzetzki, and Henrik~J Kleven.
\newblock Estimating the elasticity of intertemporal substitution using
  mortgage notches.
\newblock \emph{The Review of Economic Studies}, 87\penalty0 (2):\penalty0
  656--690, 2020.

\bibitem[Blomquist et~al.(2021)Blomquist, Newey, Kumar, and
  Liang]{blomquist2021bunching}
S{\"o}ren Blomquist, Whitney~K Newey, Anil Kumar, and Che-Yuan Liang.
\newblock On bunching and identification of the taxable income elasticity.
\newblock \emph{Journal of Political Economy}, 129\penalty0 (8):\penalty0
  2320--2343, 2021.

\bibitem[Caetano(2015)]{caetano2015test}
Carolina Caetano.
\newblock A test of exogeneity without instrumental variables in models with
  bunching.
\newblock \emph{Econometrica}, 83\penalty0 (4):\penalty0 1581--1600, 2015.

\bibitem[Caetano et~al.(2022)Caetano, Caetano, and
  Nielsen]{caetano2022identification}
Carolina Caetano, Gregorio Caetano, and Eric Nielsen.
\newblock Identification and estimation of average marginal treatment effects
  with a bunching design.
\newblock Technical report, Working paper, 2022.

\bibitem[Caetano et~al.(2024)Caetano, Caetano, and
  Nielsen]{caetano2023correcting}
Carolina Caetano, Gregorio Caetano, and Eric Nielsen.
\newblock Correcting for endogeneity in models with bunching.
\newblock \emph{Journal of Business \& Economic Statistics}, pages 1--13, 2024.

\bibitem[Caetano et~al.(2025)Caetano, Caetano, Goff, and
  Nielsen]{caetano2025identification}
Carolina Caetano, Gregorio Caetano, Leonard Goff, and Eric Nielsen.
\newblock Identification of causal effects with a bunching design.
\newblock 2025.

\bibitem[Card et~al.(2017)Card, Lee, Pei, and Weber]{card2017regression}
David Card, David~S Lee, Zhuan Pei, and Andrea Weber.
\newblock Regression kink design: Theory and practice.
\newblock In \emph{Regression discontinuity designs: Theory and applications},
  pages 341--382. Emerald Publishing Limited, 2017.

\bibitem[Chen et~al.(2021)Chen, Liu, Su{\'a}rez~Serrato, and
  Xu]{chen2021notching}
Zhao Chen, Zhikuo Liu, Juan~Carlos Su{\'a}rez~Serrato, and Daniel~Yi Xu.
\newblock Notching r\&d investment with corporate income tax cuts in china.
\newblock \emph{American Economic Review}, 111\penalty0 (7):\penalty0
  2065--2100, 2021.

\bibitem[Chetty et~al.(2011)Chetty, Friedman, Olsen, and
  Pistaferri]{chetty2011adjustment}
Raj Chetty, John~N Friedman, Tore Olsen, and Luigi Pistaferri.
\newblock Adjustment costs, firm responses, and micro vs. macro labor supply
  elasticities: Evidence from {D}anish tax records.
\newblock \emph{The Quarterly Journal of Economics}, 126\penalty0 (2):\penalty0
  749--804, 2011.

\bibitem[Cox et~al.(2021)Cox, Liu, and Morrison]{cox2021market}
Natalie Cox, Ernest Liu, and Daniel Morrison.
\newblock Market power in small business lending: A two-dimensional bunching
  approach.
\newblock Technical report, 2021.

\bibitem[Diamond and Persson(2017)]{diamond2017long}
Rebecca Diamond and Petra Persson.
\newblock The long-term consequences of teacher discretion in grading of
  high-stakes tests.
\newblock Technical report, National Bureau of Economic Research, 2017.

\bibitem[Einav et~al.(2015)Einav, Finkelstein, and Schrimpf]{einav2015response}
Liran Einav, Amy Finkelstein, and Paul Schrimpf.
\newblock The response of drug expenditure to nonlinear contract design:
  Evidence from medicare part d.
\newblock \emph{The Quarterly Journal of Economics}, 130\penalty0 (2):\penalty0
  841--899, 2015.

\bibitem[Goff(2022)]{goff2022treatment}
Leonard Goff.
\newblock Treatment effects in bunching designs: The impact of mandatory
  overtime pay on hours.
\newblock \emph{arXiv preprint arXiv:2205.10310}, 2022.

\bibitem[Kleven and Waseem(2013)]{kleven2013using}
Henrik~J Kleven and Mazhar Waseem.
\newblock Using notches to uncover optimization frictions and structural
  elasticities: Theory and evidence from pakistan.
\newblock \emph{The Quarterly Journal of Economics}, 128\penalty0 (2):\penalty0
  669--723, 2013.

\bibitem[Kleven(2016)]{kleven2016bunching}
Henrik~Jacobsen Kleven.
\newblock Bunching.
\newblock \emph{Annual Review of Economics}, 8:\penalty0 435--464, 2016.

\bibitem[Lee and Lemieux(2010)]{lee2010regression}
David~S Lee and Thomas Lemieux.
\newblock Regression discontinuity designs in economics.
\newblock \emph{Journal of Economic Literature}, 48\penalty0 (2):\penalty0
  281--355, 2010.

\bibitem[Pollinger(2025)]{pollinger2025kinks}
Stefan Pollinger.
\newblock Kinks know more: Estimating intensive and participation margin
  responses using nonlinear budget sets.
\newblock 2025.

\bibitem[Saez(2010)]{saez2010taxpayers}
Emmanuel Saez.
\newblock Do taxpayers bunch at kink points?
\newblock \emph{American Economic Journal: Economic Policy}, 2\penalty0
  (3):\penalty0 180--212, 2010.

\bibitem[Yu(2015)]{yu2015universal}
Hao Yu.
\newblock Universal health insurance coverage for 1.3 billion people: What
  accounts for {C}hina's success?
\newblock \emph{Health Policy}, 119\penalty0 (9):\penalty0 1145--1152, 2015.
\newblock \doi{10.1016/j.healthpol.2015.07.005}.

\end{thebibliography}
\endgroup

\clearpage
\begin{appendices}

\begin{center}
\huge{\textbf{ONLINE APPENDIX}}
\end{center}

\appendix
\setcounter{table}{0}
\renewcommand{\thetable}{A\arabic{table}}

\setcounter{figure}{0}
\renewcommand{\thefigure}{A\arabic{figure}}

\pagenumbering{arabic} 
\setcounter{page}{1}   



\section{Additional Tables and Figures}
\label{app:additional}

\begin{figure}[htbp]
\centering
\caption{\small{Observed and counterfactual last visit  quarter, URRBMI 2012 (baseline specification).}}
\includegraphics[width=0.45\linewidth]{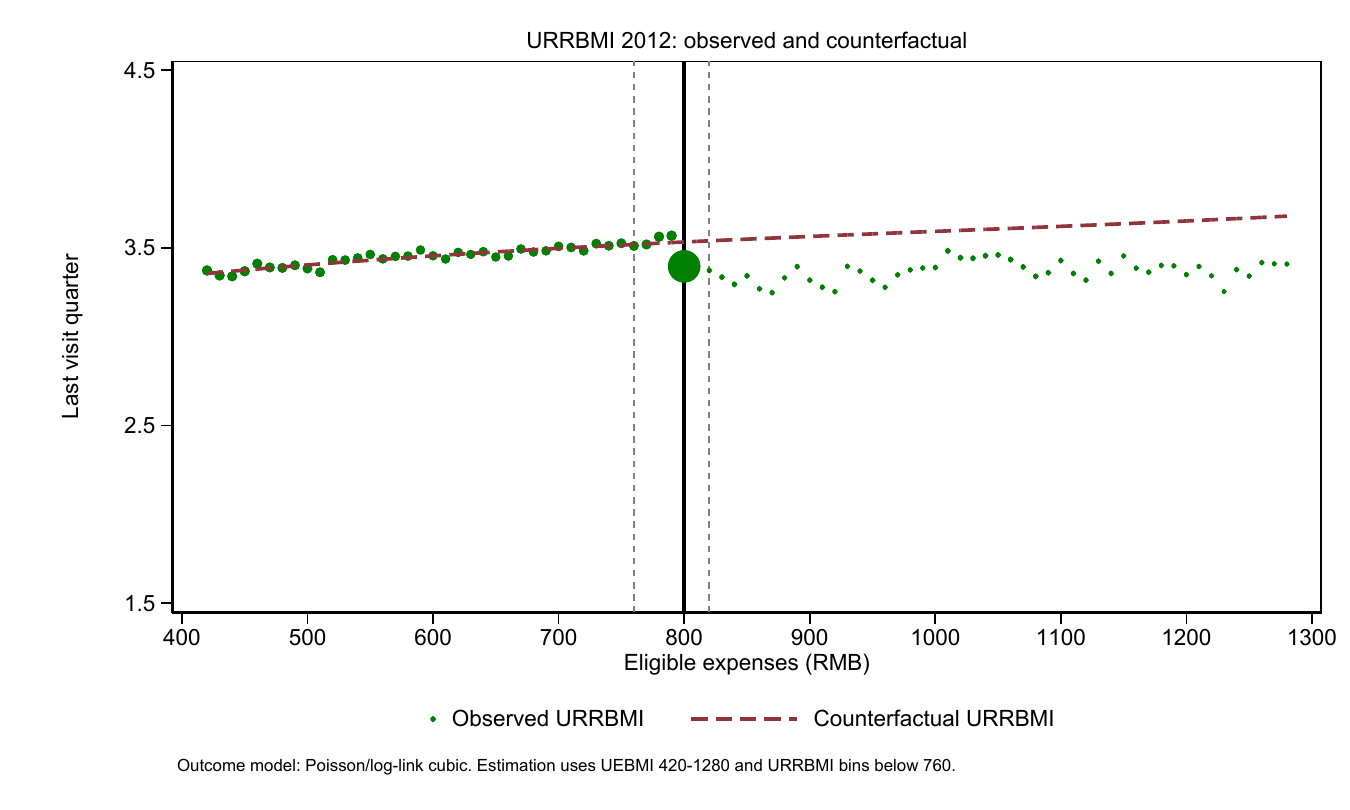}
\begin{minipage}{16cm}
\footnotesize{Notes: Green markers are bin-level mean last visit  quarter  for observed URRBMI; the dashed line is the design-based counterfactual $m_F^{cf}$. Reference lines and estimation windows as in Figure~\ref{fig:app-cf-hosp-share}. Above the cap the observed last quarter falls well below the counterfactual, consistent with enrollees stopping care once their reimbursement is exhausted. Outcome model Poisson/log-link cubic.}
\end{minipage}
\label{fig:app-cf-last-m}
\end{figure}
\begin{figure}[htbp]
\centering
\caption{\small{Observed and counterfactual hospital visit share, URRBMI 2012 (baseline specification).}}
\includegraphics[width=0.45\linewidth]{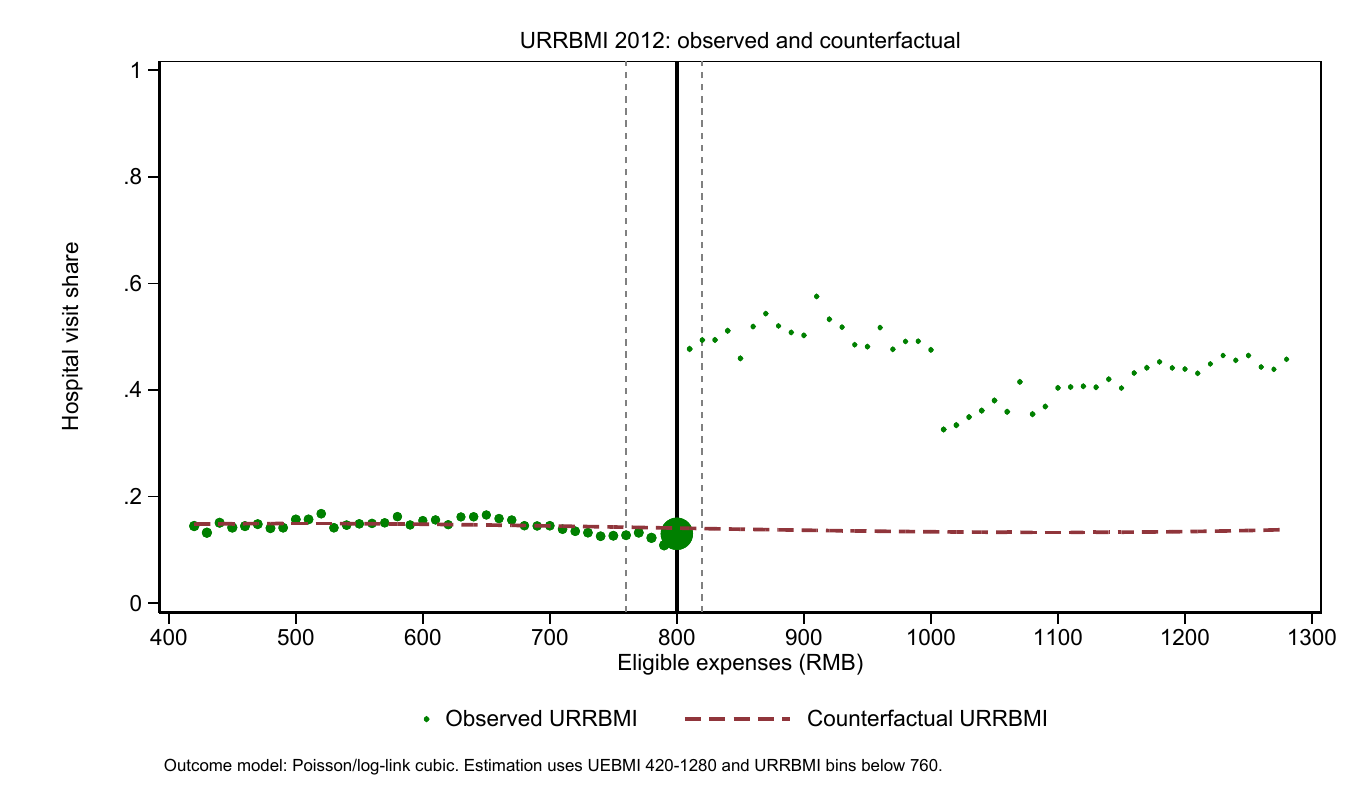}
\begin{minipage}{16cm}
\footnotesize{Notes: Green markers are bin-level mean hospital visit share for observed URRBMI; the dashed line is the design-based counterfactual $m_F^{cf}$, identified from the UEBMI profile and the adjustment $\delta_y$ fit on URRBMI bins below the cutoff. The solid vertical marks the kink $x^*=800$~RMB; inner dashed verticals mark the diffuse window $[760,820]$. Outcome model Poisson/log-link cubic; estimation uses UEBMI $[420,1280]$ and URRBMI bins below $760$.}
\end{minipage}
\label{fig:app-cf-hosp-share}
\end{figure}

This appendix reports the observed and design-based counterfactual profiles for the secondary outcomes summarized in Table~\ref{tab:app-effects}: the timing of the last visit (Figure~\ref{fig:app-cf-last-m}), the composition of care---the hospital share of visits (Figure~\ref{fig:app-cf-hosp-share}). 

\end{appendices}

\end{document}